\documentclass[11pt,a4paper]{article}
\usepackage[margin=25mm]{geometry}
\usepackage[T1]{fontenc}
\usepackage[utf8]{inputenc}
\usepackage[english]{babel}
\usepackage{graphicx}
\usepackage{amsmath,amssymb}
\usepackage{booktabs}
\usepackage{placeins}

\newcommand{\RRthreepanels}[3]{%
  \begingroup
  \setlength{\tabcolsep}{1pt}%
  \begin{tabular}{@{}ccc@{}}
    \includegraphics[width=150pt]{#1} &
    \includegraphics[width=150pt]{#2} &
    \includegraphics[width=150pt]{#3}
    \\[-2pt]
    a & b & c
  \end{tabular}%
  \endgroup
}

\newcommand{\RRverticalpanels}[3]{%
  \begingroup
  \setlength{\tabcolsep}{2pt}%
  \begin{tabular}{@{}cc@{}}
    \includegraphics[width=360pt]{#1} & a
    \\[-5pt]
    \includegraphics[width=360pt]{#2} & b
    \\[-5pt]
    \includegraphics[width=360pt]{#3} & c
  \end{tabular}%
  \endgroup
}

\newcommand{\RRprofileverticalpanels}[3]{%
  \begingroup
  \begin{tabular}{@{}c@{}}
    \includegraphics[width=225pt]{#1}\\[-2pt]
    a\\[7pt]
    \includegraphics[width=225pt]{#2}\\[-2pt]
    b\\[7pt]
    \includegraphics[width=225pt]{#3}\\[-2pt]
    c
  \end{tabular}%
  \endgroup
}


\usepackage[numbers,sort&compress]{natbib}
\title{Regular reflection of shock waves: non-equilibrium effects for various steady flow parameters}
\author{Yevgeniy Bondar\thanks{ \texttt{bond@itam.nsc.ru}}\\
  \small Faculty of Physics, Lomonosov Moscow State University, Moscow, 119991, Russia\\
  \small Khristianovich Institute of Theoretical and Applied Mechanics SB RAS,\\[-2pt]
  \small Novosibirsk, 630090, Russia
  \and
  Maksim Timokhin\thanks{Corresponding author: \texttt{timokhin@physics.msu.ru}}\\
  \small Faculty of Physics, Lomonosov Moscow State University, Moscow, 119991, Russia\\
  \small Khristianovich Institute of Theoretical and Applied Mechanics SB RAS,\\[-2pt]
  \small Novosibirsk, 630090, Russia}
\date{}
\begin{document}
\maketitle
\begin{abstract}
A parametric study of the non-equilibrium region formed by the steady regular reflection of two symmetric oblique shock waves in a monatomic gas is performed using the direct simulation Monte Carlo (DSMC) method. Three series of calculations are considered, with the incident-shock normal Mach number fixed at 2, 4 and 8, respectively. Within each series, the incidence angle is reduced while the normal strength of the incident shock remains unchanged. The decrease in angle primarily results in a longer non-equilibrium region of shock-wave interaction. Over the parameter range considered, the density-transition width along the plane of symmetry is approximately inversely proportional to the sine of the incidence angle. Increasing the normal Mach number enhances temperature anisotropy, the local temperature overshoot relative to the equilibrium downstream state, and the streamwise-velocity undershoot. 
\end{abstract}
\noindent\textbf{Keywords:} regular shock-wave reflection, DSMC, kinetic non-equilibrium, temperature anisotropy
\par\medskip

\section{Introduction}
\label{sec_intro}

Shock-wave interactions are among the fundamental phenomena that determine the structure of supersonic flows~\cite{BenDor2007}.  Among the different interaction configurations, two-dimensional regular reflection (RR) is of particular interest. In the inviscid approximation, the incident oblique shock (IS) and reflected shock (RS) meet at a single point on the plane of symmetry.
In the viscous approximation, it provides a sensitive test of models for viscosity, heat conduction and kinetic non-equilibrium~\cite{BondarShoevTimokhin2024}.

A typical symmetric RR configuration is produced by a supersonic flow over two wedges arranged symmetrically about the flow axis. The incident oblique shocks generated by the wedges interact at the plane of symmetry, giving rise to two reflected shocks. A schematic of the configuration is shown in figure~\ref{fig_statement}. 
In the present work
the focus is on the local shock-interaction region, sufficiently far from the downstream influence of the wedge trailing edges.
Hereafter, $\alpha$ denotes the angle between the incident shock front and the plane of symmetry. For an oblique shock to exist, this angle must exceed the Mach angle, $\alpha_{\mathrm{M}}=\arcsin(1/M_\infty)$, where $M_\infty$ is the free-stream Mach number. The position of the selected flow conditions relative to the existence boundary for a weak inviscid RR solution is assessed using the detachment criterion. At this boundary, the reflected shock on the weak branch of the shock polar can no longer turn the flow back to the direction of the plane of symmetry \cite{BenDor2007,Hornung1986}. Finite shock thickness can also affect the transition from regular to Mach reflection. Shoev et al.~\cite{ShoevNazarovTimokhin2025} demonstrated that viscous and kinetic-model calculations may retain a two-shock configuration under conditions where inviscid shock-polar theory predicts Mach reflection. In contrast, the present cases are selected well within the inviscid regular-reflection domain, allowing us to focus on the internal non-equilibrium structure rather than on the reflection transition.

When viscosity and heat conduction are taken into account, the flow near the RR point develops an essentially two-dimensional structure. An extended non-equilibrium region forms along the plane of symmetry, where the flow parameters remain different from the equilibrium state predicted by the inviscid Rankine--Hugoniot (RH) relations over distances of tens of mean free paths \cite{Khotyanovsky2009,BondarShoevTimokhin2024}. Related non-Rankine--Hugoniot regions have also been investigated in irregular and Mach reflection \cite{ste59,Ivanov2010ProgresAero,jfm_2023}.

An important related class of shock interactions involves centreline reflection in axisymmetric flows. Shoev and Ogawa \cite{ShoevOgawa2019} demonstrated that viscous effects can substantially alter the reflection configuration, producing a two-shock structure at low Reynolds numbers instead of the three-shock configuration predicted by inviscid theory. Hew et al.\ \cite{HewOgawaBoswell2024} subsequently investigated centreline shock reflection in overexpanded supersonic microjets and showed that viscosity and rarefaction effects can similarly influence the transition between two- and three-shock configurations. More recently, Higa et al.\ \cite{HigaEtAl2025} investigated the non-Rankine--Hugoniot zone in axisymmetric supersonic flow using experimental measurements and numerical simulations. Although these axisymmetric configurations differ from planar RR, the results demonstrate the importance of finite shock thickness and multidimensional transport in determining the structure of shock-interaction regions.

Recent studies have also addressed the internal non-equilibrium structure of triple points in the irregular reflection. Using a diatomic 10-moment model, Morin et al.\ \cite{MorinEtAl2026} investigated triple-point interactions and identified an internal connecting shock associated with a localized translational-temperature enhancement and pronounced temperature anisotropy. Although their study concerns triple-point configurations rather than the two-shock structure of RR, it further demonstrates that the internal structure of shock intersections may exhibit non-equilibrium features that cannot be inferred from inviscid jump relations alone.

The distributions of macroscopic flow parameters along the plane of symmetry have several qualitative similarities to the internal structure of a planar shock wave. Both involve a continuous transition between two equilibrium states accompanied by pronounced non-equilibrium of the translational degrees of freedom. A temperature overshoot is known to occur in strong planar shock waves \cite{Erofeev_Friedlander_overshoot_2002, overshoot_2015}, whereas in RR, viscosity and two-dimensional transport can produce local temperature and pressure maxima near the reflection point \cite{Khotyanovsky2009,BondarShoevTimokhin2024}. Recent discrete Boltzmann calculations of planar shock structures in argon have also provided detailed information on higher-order non-equilibrium moments and their dependence on shock intensity, with comparisons against experimental and DSMC results \cite{ZhangGanYang2026}.

Non-equilibrium effects in shock-wave interactions have also been studied using mesoscopic kinetic approaches. Qiu et al.\ \cite{QiuEtAl2020} investigated the formation and curvature of the shock-wave fronts in RR and the influence of viscosity on the non-equilibrium structure. A broader review of mesoscopic approaches to non-equilibrium effects in shock waves and related interaction phenomena was subsequently presented in \cite{QiuYangBao2024}. A more recent study based on a discrete Boltzmann model examined thermodynamic non-equilibrium and entropy production during RR, including their dependence on the Mach number \cite{GanEtAl2025}.

The local molecular velocity distribution functions within the reflection zone exhibit a pronounced multimodal structure \cite{TimokhinBondarMath2025}. This finding motivated an extension of the Mott-Smith method, originally developed for the structure of a planar shock wave \cite{Mott-Smith_1951}, to the two-dimensional RR region \cite{TimokhinKudryavtsevBondar2022,TimokhinBondarLOJM2025}. It has also been shown that the two-dimensional transport of momentum and energy near RR differs substantially from transport in a planar shock wave. In particular, a streamwise-velocity minimum, distinctive distributions of viscous stress components, and pronounced transverse energy fluxes appear \cite{BondarShoevTimokhin2024}.

The previous detailed study of these effects considered a single strong RR case \cite{BondarShoevTimokhin2024}. The influence of the incident-shock angle and its normal Mach number on the spatial extent and kinetic structure of the non-equilibrium region therefore remains insufficiently characterized. Despite the available results, the effects of incident-shock intensity and interaction geometry are generally coupled. Changing the free-stream Mach number or wedge angle simultaneously alters the normal shock strength, shock orientation, flow-deflection angle and tangential velocity component. Studying isolated flow conditions therefore does not allow one to distinguish reliably between features primarily associated with incident-shock intensity and those associated with the geometry of the interaction.

To trace these trends separately, we consider series with a fixed incident-shock normal Mach number, $M_n=M_\infty\sin\alpha$. For a given $M_n$, the thermodynamic jumps across the incident shock remain unchanged, while varying $M_\infty$ changes the incidence angle $\alpha$ and the associated tangential flow component. Comparing cases within one series thus characterizes the effects of this geometric variation at a fixed incident-shock normal strength; comparing different $M_n$ series reveals the effect of changing incident-shock strength. Fixing $M_n$, however, does not imply that the Mach number ahead of the reflected shock or the reflected-shock strength is constant.

The aim of the present work is a parametric investigation of the geometry and kinetic structure of the non-equilibrium region in steady RR. Three series with $M_n=2$, $4$ and $8$ are studied; within each series, $\alpha$ is progressively decreased while $M_n=M_\infty\sin\alpha$ remains fixed. The DSMC results are used to analyse two-dimensional density and translational-temperature fields, as well as distributions of density, temperature, streamwise velocity, directional temperatures and total enthalpy along the plane of symmetry. Particular attention is paid to the quantitative angular scaling of the density-transition region width along the plane of symmetry
and the 
amplitudes of local non-equilibrium effects.

\section{Problem formulation} \label{sec_statement}
\subsection{Steady regular-reflection problem} \label{subsec_2D_statement}

\begin{figure}
\centering
    {\includegraphics[width=270pt]{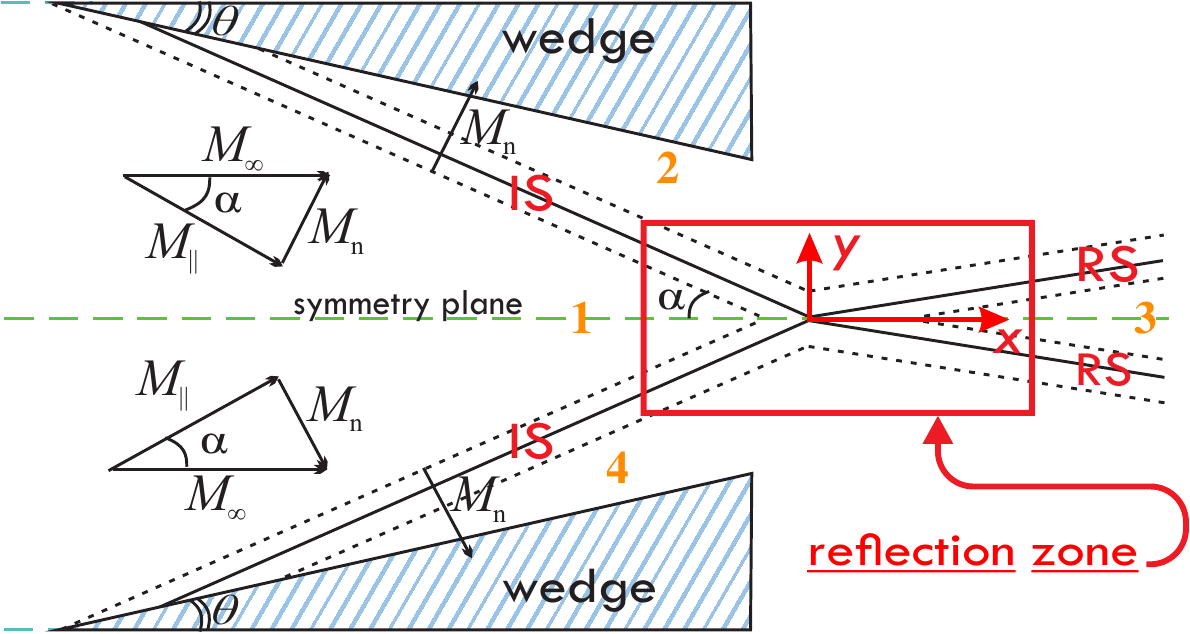}}
\caption{Schematic of steady regular shock-wave reflection.
}
\label{fig_statement}
\end{figure}

The steady RR problem is considered in a formulation similar to that of \cite{BondarShoevTimokhin2024}. Two incident shocks are generated by two wedges placed in a supersonic viscous flow. The flow configuration is shown in figure~\ref{fig_statement}. Dashed lines schematically indicate the finite-thickness shock-wave fronts upstream and downstream of the interaction region, whereas solid lines represent the positions of the discontinuities in the corresponding inviscid solution.

The equilibrium states of the corresponding inviscid solution are numbered in figure~\ref{fig_statement} and used as reference states throughout the paper. State 1 denotes the undisturbed free stream, while states 2 and 4 correspond to the regions behind the upper and lower incident shocks, respectively, and ahead of the reflected shocks. State 3 represents the equilibrium flow downstream of the reflected shocks. Owing to the symmetry of the configuration, states 2 and 4 have identical thermodynamic properties, while their transverse velocity components are equal in magnitude and opposite in sign. In the DSMC solutions, the discontinuities of the inviscid model are replaced by finite-thickness shock-wave fronts, and the flow within the interaction region exhibits pronounced kinetic non-equilibrium.

For the monatomic gas considered here, the specific-heat ratio is $\gamma=5/3$. The free-stream Mach number is defined as
\begin{equation}
M_\infty =
\frac{v_\infty}{\sqrt{\gamma RT_\infty}}
=
\frac{v_\infty}{\sqrt{5RT_\infty/3}},
\end{equation}
where $v_\infty$ and $T_\infty$ are the free-stream velocity and temperature, respectively. The free-stream Mach number and the wedge angle $\theta$ are selected so that a weak regular-reflection solution exists and all considered cases remain below the inviscid detachment boundary. The angles of the IS and RS, as well as the equilibrium flow parameters behind them, can be determined analytically.

Three series of RR calculations are considered, corresponding to fixed values of the incident-shock normal Mach number $M_n=2$, $4$ and $8$. Within each series, $M_n=M_\infty\sin\alpha$ is kept constant, while the incidence angle $\alpha$ is reduced by increasing the free-stream Mach number $M_\infty$. This choice makes it possible to investigate changes in the RR structure as $\alpha$ decreases without changing the normal strength of the incident shock. The parameters of the selected cases are listed in table~\ref{tab_cases}.

\subsubsection{Inviscid solutions for the selected cases: shock polars}
\label{subsubsec_polars}

The chosen parameterization allows us to compare trends associated with changes in the incident-shock normal strength with those observed when $\alpha$ is varied within series at fixed $M_n$.

At fixed $M_n$, the pressure, density and temperature ratios across the incident shock, $p_2/p_1$, $\rho_2/\rho_1$ and $T_2/T_1$, remain unchanged \cite{Rankine}. In particular,
\begin{equation}
    p_{12}
    \equiv
    \frac{p_2}{p_1}
    =
    \frac{2\gamma M_n^2-(\gamma-1)}{\gamma+1}.
    \label{eq_p12_Mn}
\end{equation}
However, the flow direction behind the incident shock and the Mach number $M_2$ depend on $M_\infty$ and $\alpha$.

As $\alpha$ decreases, the component of the free-stream Mach number tangential to the incident-shock front increases:
\begin{equation}
    M_{\parallel}
    =
    M_\infty\cos\alpha
    =
    \sqrt{M_\infty^2-M_n^2}.
    \label{eq_Mparallel}
\end{equation}
The normal component of the downstream Mach number immediately behind the incident shock is given by
\begin{equation}
    M_{n2}^{\,2}
    =
    \frac{1+\dfrac{\gamma-1}{2}M_n^2}
    {\gamma M_n^2-\dfrac{\gamma-1}{2}},
\end{equation}
and the total Mach number in state 2 is
\begin{equation}
    M_2
    =
    \frac{M_{n2}}{\sin(\alpha-\theta)}.
    \label{eq_M2}
\end{equation}
Thus, the cases within each series have the same incident-shock normal strength but differ appreciably in the tangential component of the free-stream velocity, the flow direction behind the IS, and the total Mach number $M_2$ ahead of the RS.

\begin{figure}
\centering
\begingroup
\setlength{\tabcolsep}{1pt}
\begin{tabular}{@{}ccc@{}}
{\includegraphics[width=150pt]{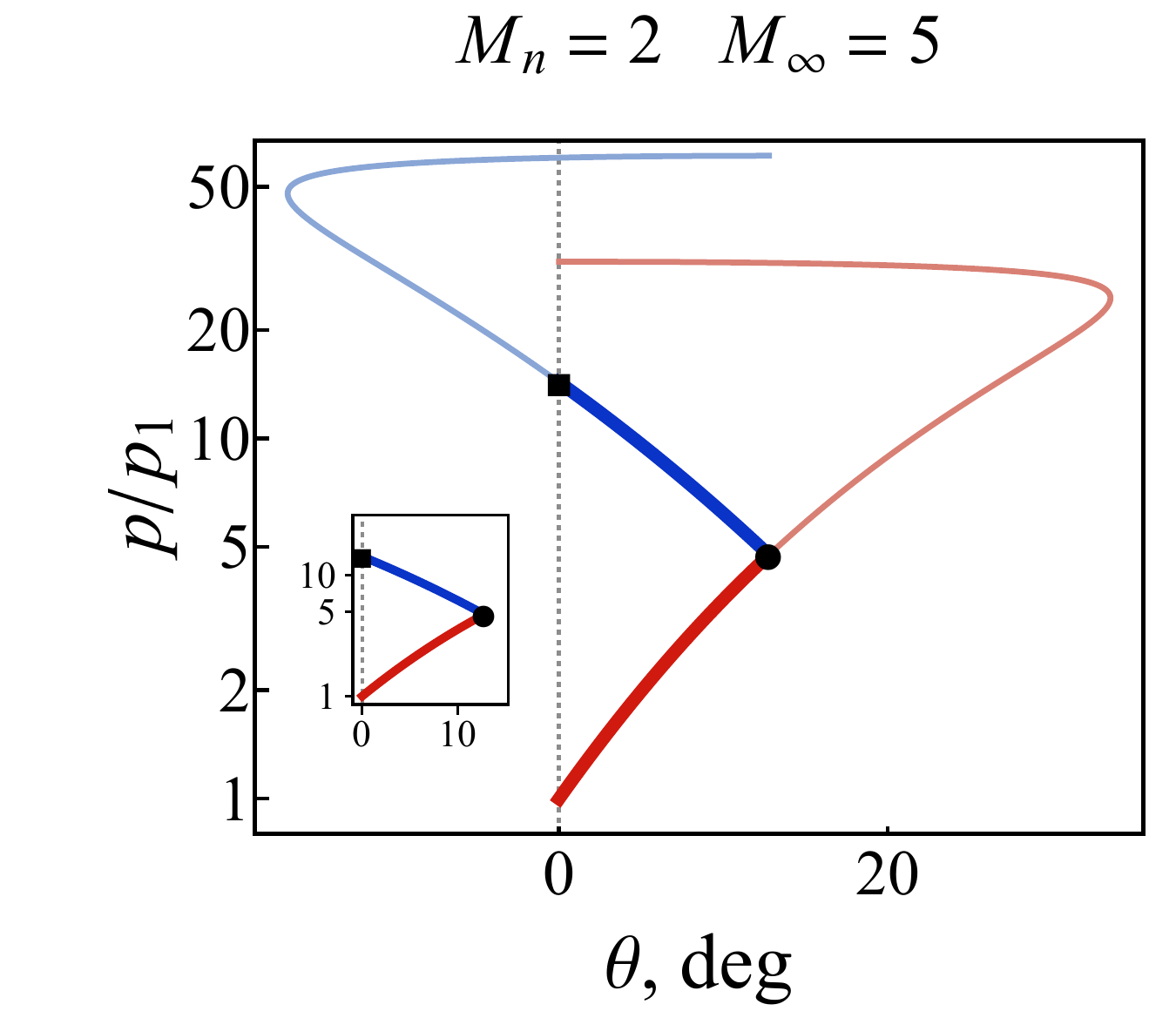}}&
{\includegraphics[width=150pt]{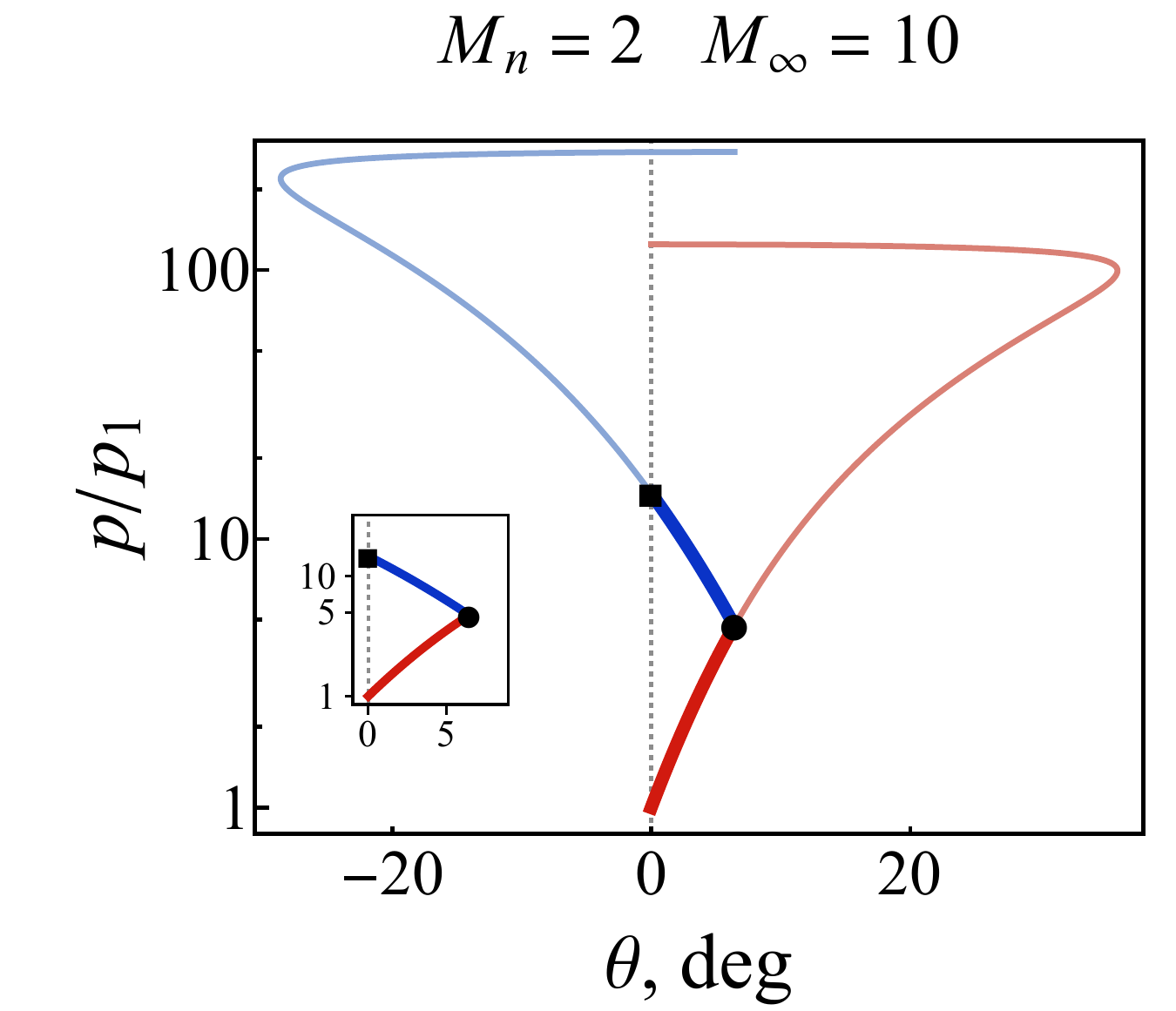}}&
{\includegraphics[width=150pt]{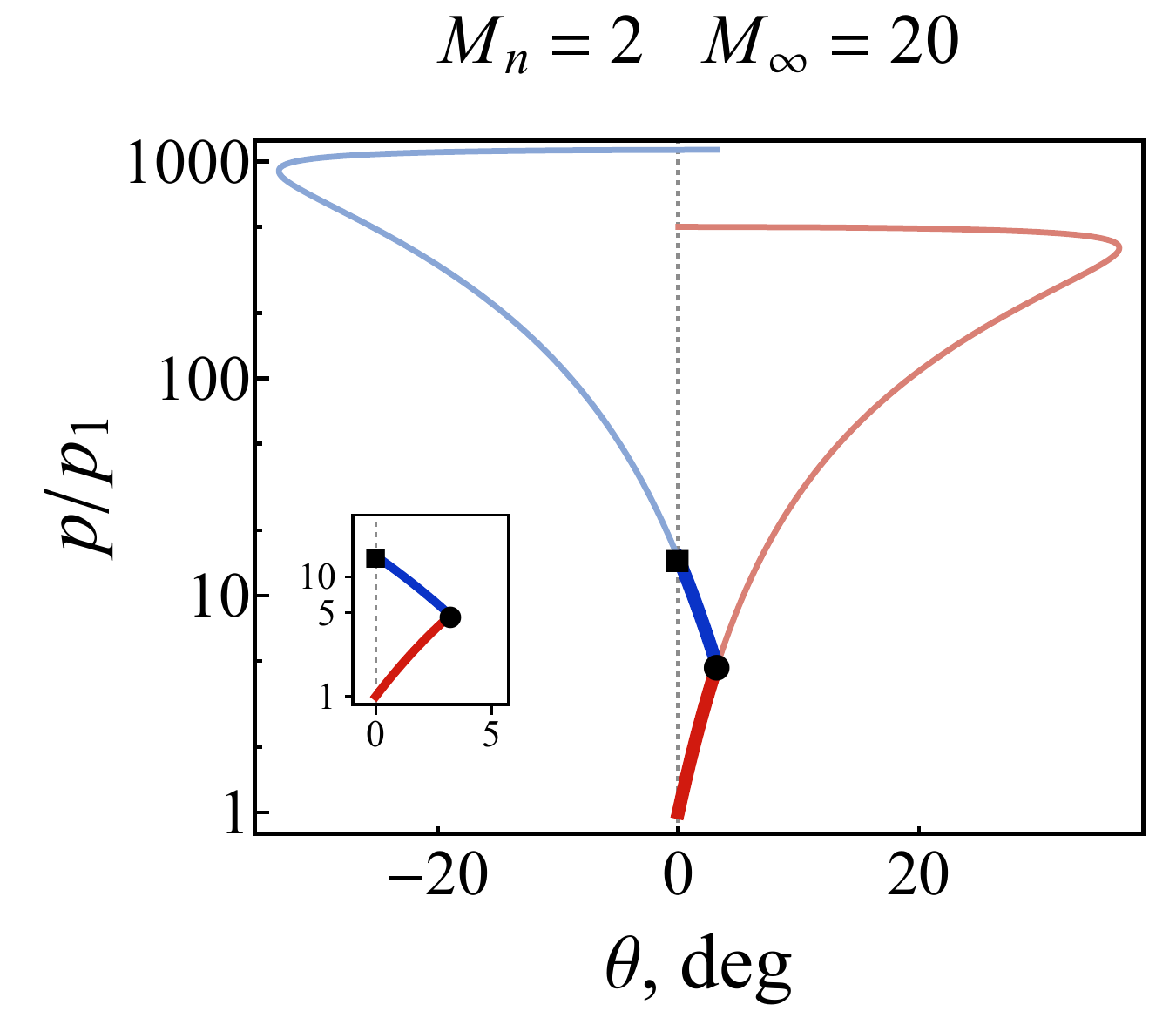}}
\\
a&b&c
\\
{\includegraphics[width=150pt]{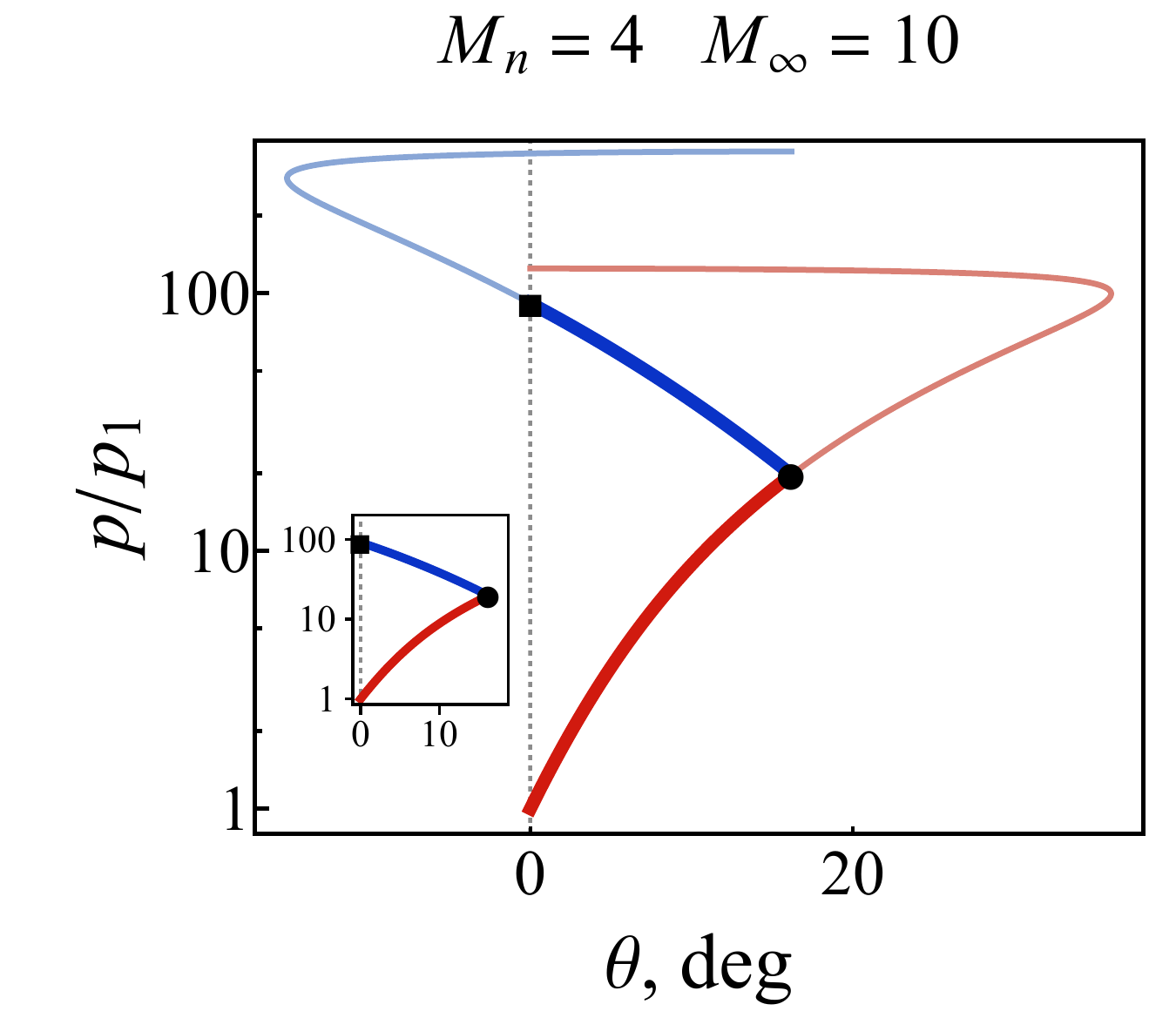}}&
{\includegraphics[width=150pt]{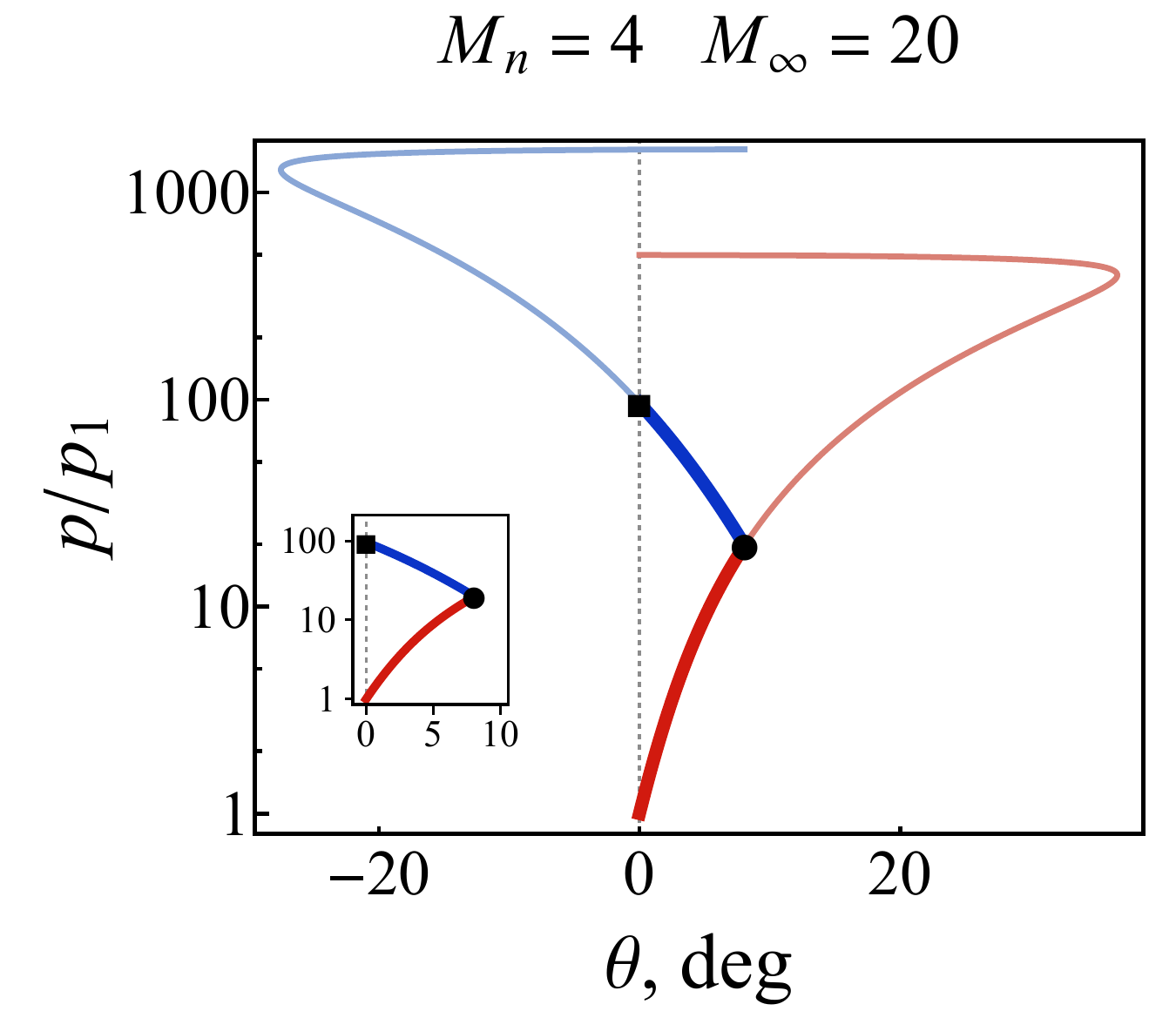}}&
{\includegraphics[width=150pt]{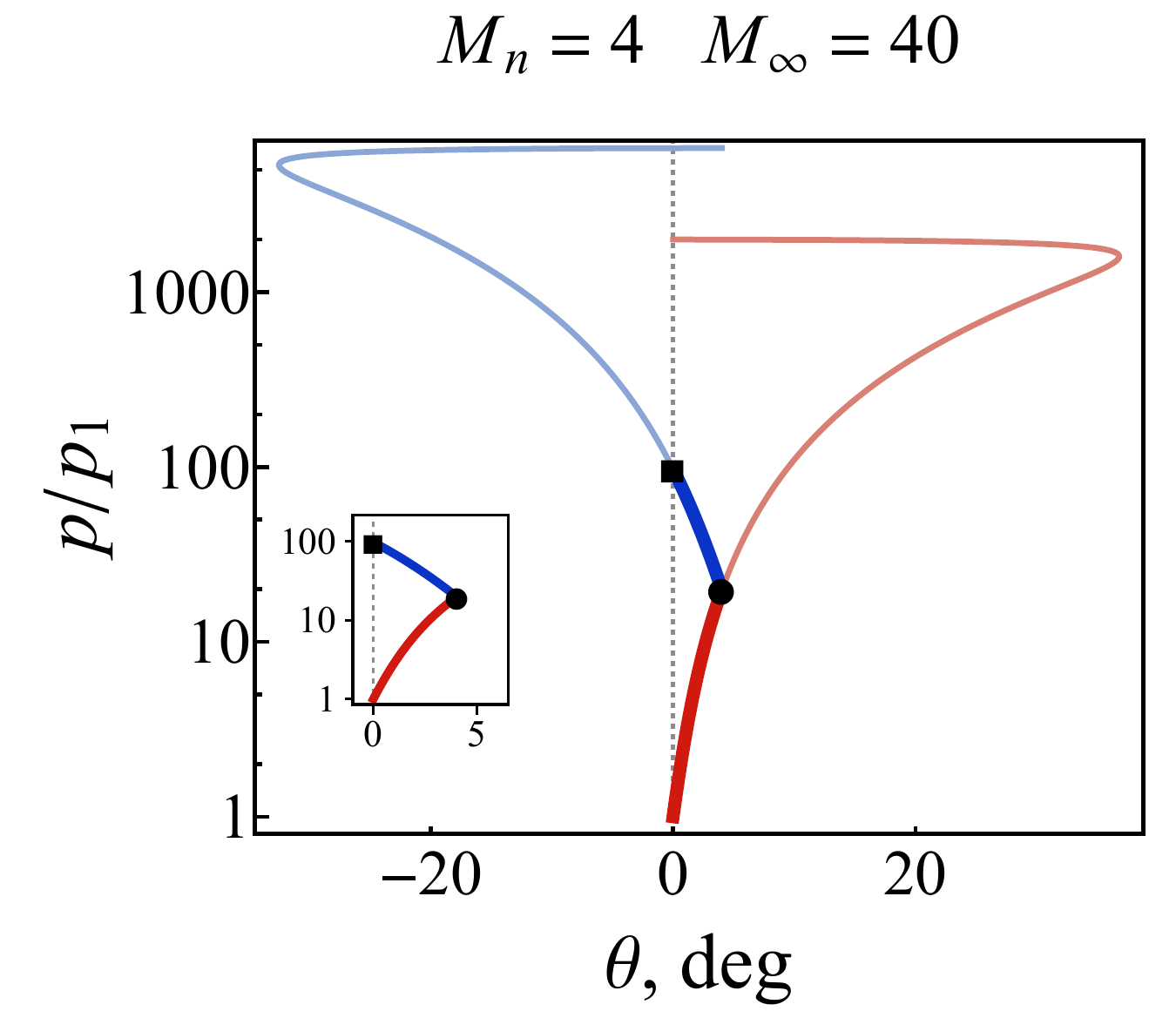}}
\\
d&e&f
\\
{\includegraphics[width=150pt]{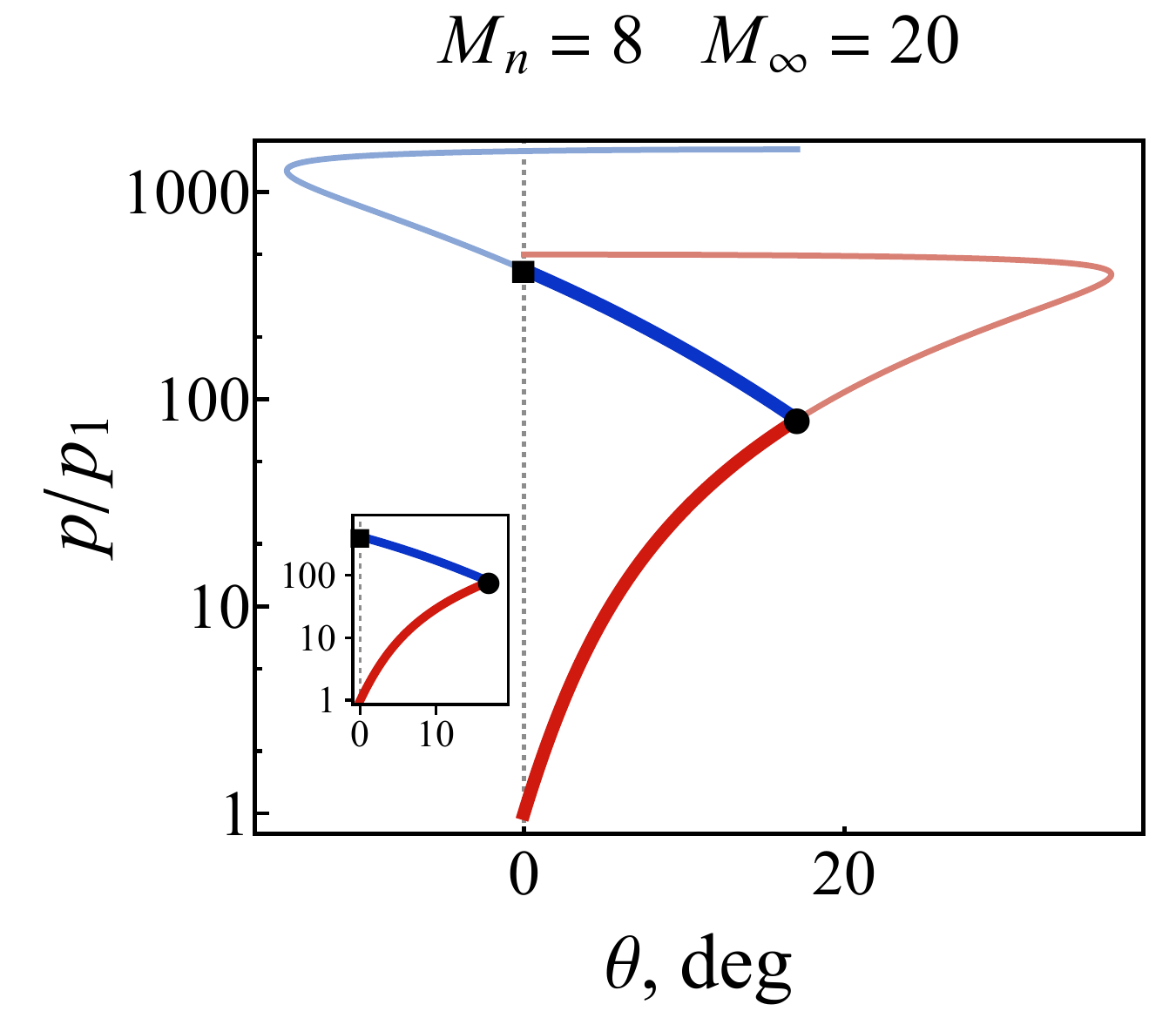}}&
{\includegraphics[width=150pt]{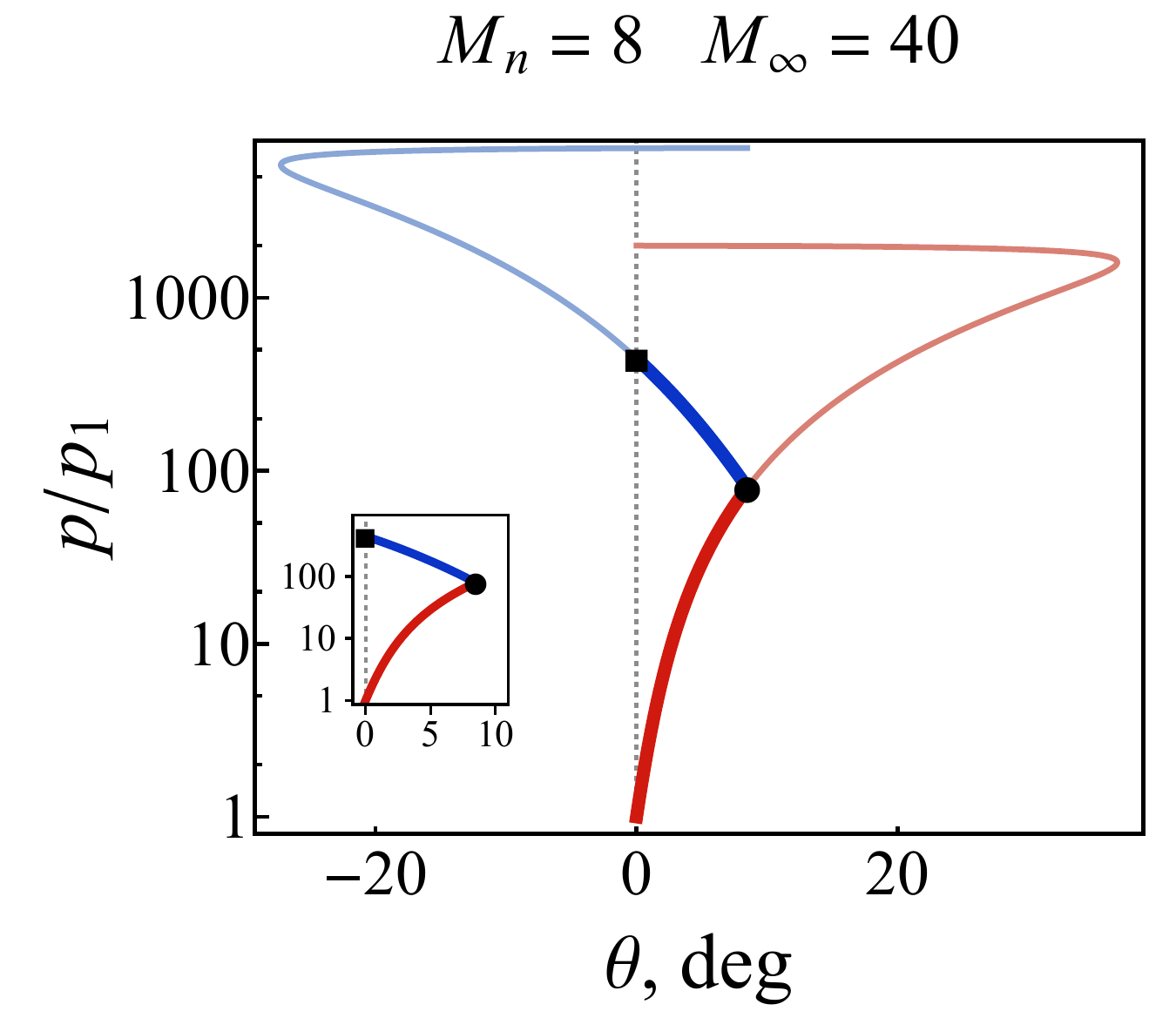}}&
{\includegraphics[width=150pt]{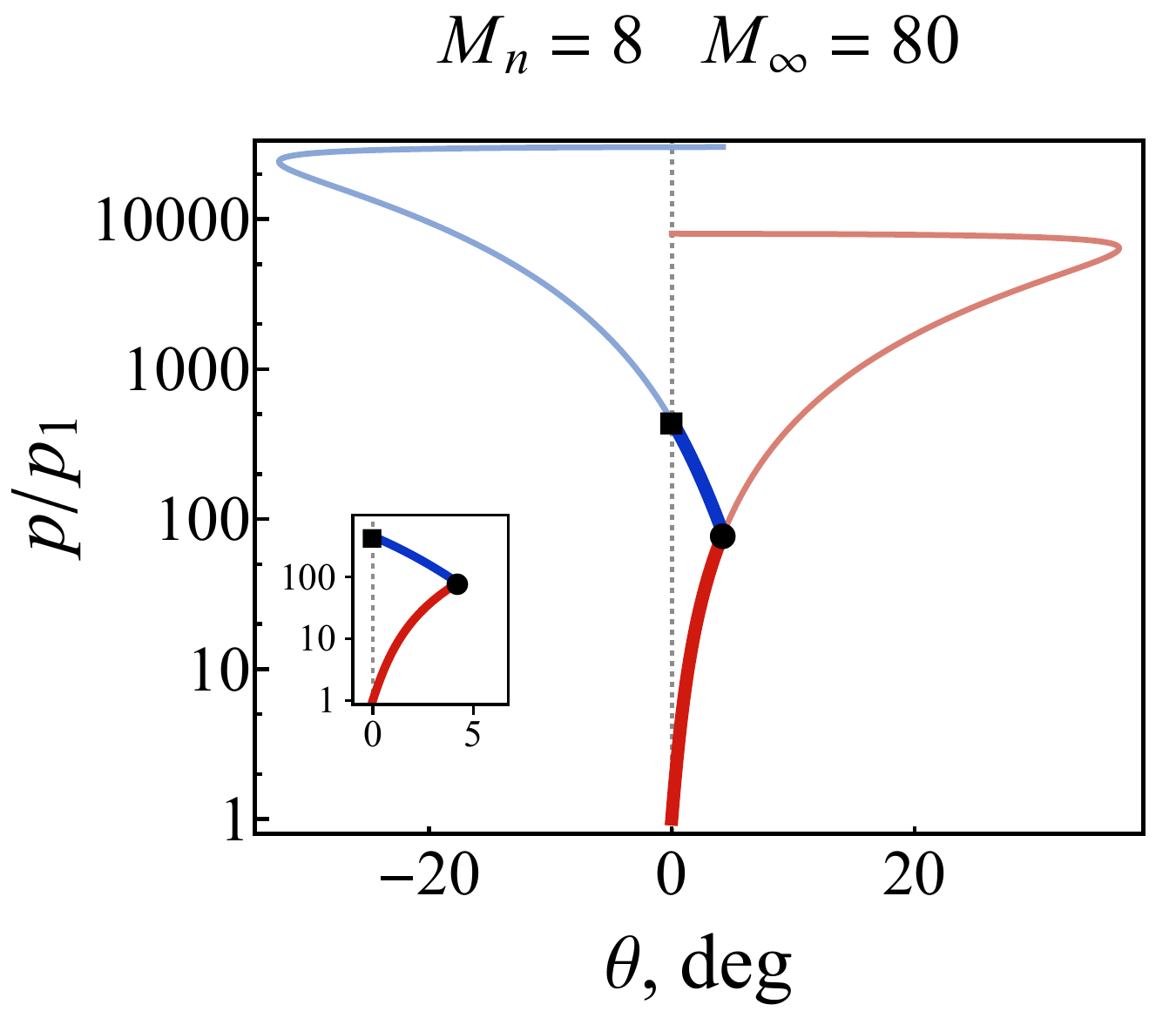}}
\\
g&h&i
\end{tabular}
\endgroup
\caption{Shock polars for all the cases in table~\ref{tab_cases}. Red and blue curves denote the incident and reflected shock polars, respectively. Thick segments indicate the selected weak RR transitions, while circles and squares mark equilibrium states 2 and 3. Insets show enlarged views of the selected branches.}
\label{fig_polars}
\end{figure}

The shock polars for the incident and reflected shocks in all the selected cases are shown in figure~\ref{fig_polars}. State 1 represents the undisturbed free stream, with $\theta_1=0$ and $p/p_1=1$, whereas state 2 behind the upper incident shock is characterized by $\theta_2=\theta$ and $p_2/p_1=p_{12}$. Here, $\theta$ denotes both the flow-deflection angle across the incident shock and the wedge angle.

The reflected-shock polar is first constructed using the flow direction in state 2 as the angular reference. Let $\delta\geq0$ denote the magnitude of the flow deflection across the reflected shock towards the plane of symmetry. Defining
\begin{equation}
    p_{23}
    \equiv
    \frac{p_3}{p_2},
\end{equation}
the transformation to the global coordinates gives
\begin{equation}
    \theta_3=\theta_2-\delta,
    \qquad
    \frac{p_3}{p_1}
    =
    \frac{p_2}{p_1}
    \frac{p_3}{p_2}
    =
    p_{12}p_{23}.
    \label{eq_reflected_polar_transform}
\end{equation}

The non-permeability condition on the plane of symmetry requires $\theta_3=0$. Therefore, the reflected shock must turn the flow through an angle equal in magnitude and opposite in direction to the deflection across the incident shock.

The complete shock polar consists of weak and strong branches that meet at the maximum flow-deflection angle. In the present study, the weak branches are used to construct the inviscid reference solution, because the weak solution is observed in experiments and simulations in cases when the weak and strong solutions are both theoretically possible~\cite{LANDAU1987}.  In figure~\ref{fig_polars}, the selected transition $1\rightarrow2$ is shown as a thickened segment of the incident-shock polar, while the transition $2\rightarrow3$ is shown as a thickened segment of the reflected-shock polar. States 2 and 3 are marked by circles and squares, respectively. The strong branches are displayed only to illustrate the complete shock-polar structure and are not used in the subsequent analysis.

Figure~\ref{fig_polars} demonstrates two distinct trends in the inviscid solution. Moving from left to right along each row, $M_n$ remains constant while $M_\infty$ increases and $\alpha$ decreases. As $M_\infty$ increases, the complete incident-shock polar spans a wider range of deflection angles and pressure ratios. The selected incident shock, however, does not become stronger: the pressure, density and temperature ratios across the IS remain unchanged, while state 2 shifts towards smaller deflection angles $\theta_2$.

The conditions for the formation of the reflected shock change substantially. As $\alpha$ decreases within a series, the Mach number $M_2$ ahead of the RS increases rapidly, while the required flow deflection $\delta=\theta_2$ decreases. For example, in the $M_n=4$ series, $M_2$ increases from $3.82$ to $16.44$, whereas the required deflection decreases from $16.196^\circ$ to $4.030^\circ$. 

Moving from top to bottom within each column, $\alpha$ remains constant while $M_n$ increases from $2$ to $8$. Unlike variations along a row, the most pronounced changes are then observed in the pressure levels: states 2 and 3 shift towards substantially higher values of $p/p_1$. The deflection angle $\theta_2$ changes much less, with a tendency towards saturation that is particularly evident between $M_n=4$ and $M_n=8$. Thus, the figure distinguishes the angular rearrangement of the weak RR solution as $\alpha$ decreases within a series from the increase in pressure jumps as $M_n$ grows.
For every selected case, the weak branch of the reflected-shock polar intersects the line $\theta=0$. Therefore, according to the detachment criterion, a weak inviscid RR solution exists for all the cases considered.

\begin{table}[ht]
\centering
\caption{Shock and wedge angles for the selected cases.}
\label{tab_cases}
\begin{tabular}{cccccc}
\toprule
$M_n$ & $M_\infty$ & $\alpha,^\circ$ & $\theta,^\circ$
& $M_2$ 
\\
\midrule
2.0 &  5.0 & 23.578 & 12.768 &  3.236  
\\
2.0 & 10.0 & 11.537 &  6.434 &  6.824 
\\
2.0 & 20.0 &  5.739 &  3.221 & 13.818  
\\
\midrule
4.0 & 10.0 & 23.578 & 16.196 &  3.817  
\\
4.0 & 20.0 & 11.537 &  8.069 &  8.108 
\\
4.0 & 40.0 &  5.739 &  4.030 & 16.444  
\\
\midrule
8.0 & 20.0 & 23.578 & 17.062 &  4.038  
\\
8.0 & 40.0 & 11.537 &  8.479 &  8.591 
\\
8.0 & 80.0 &  5.739 &  4.232 & 17.429 
\\
\bottomrule
\end{tabular}
\end{table}

At fixed $M_n$, the pressure ratio across the incident shock remains constant:
\begin{equation}
    p_{12}
    =
    \begin{cases}
        4.75,  & M_n=2,\\
        19.75, & M_n=4,\\
        79.75, & M_n=8.
    \end{cases}
\end{equation}
Thus, all three cases within a series have the same value of $p_2/p_1$, corresponding to the vertical coordinate of state 2 on the shock polar, whereas its horizontal coordinate $\theta_2$ decreases with $\alpha$.

The equilibrium properties of state 3 vary relatively little within each series. In particular, the downstream pressure ratio $p_3/p_1$ lies within the following ranges:
\begin{align}
    14.33 &\leq \frac{p_3}{p_1} \leq 14.88,
    && M_n=2,\\
    91.41 &\leq \frac{p_3}{p_1} \leq 97.26,
    && M_n=4,\\
    422.08 &\leq \frac{p_3}{p_1} \leq 452.41,
    && M_n=8.
\end{align}
The relative variation of $p_3/p_1$ within the three series is approximately $3.8\%$, $6.4\%$ and $7.2\%$, respectively.

The shock polars therefore reveal two distinct trends in the evolution of the inviscid RR solution. At fixed $M_n$, decreasing $\alpha$ is accompanied by a rapid increase in $M_2$ and a reduction in the required reflected-shock deflection,
while the final pressure levels change relatively little. Conversely, increasing $M_n$ at fixed $\alpha$ primarily increases the pressure jumps across the incident and reflected shocks.

\subsubsection{Characteristic length scales and Knudsen number}
\label{subsubsec_Knudsen}

The global Knudsen number is defined as the ratio of the free-stream mean free path $\lambda_\infty$ to the length of the windward side of the wedge $w$:
\begin{equation}
    Kn = \frac{\lambda_\infty}{w}.
\end{equation}
In all the calculations presented here, the global Knudsen number is fixed at $Kn=0.001$, corresponding to $w=1000\,\lambda_\infty$.

The variable hard sphere (VHS) molecular model \cite{Bird_book} is employed to describe molecular collisions. In this model, the temperature dependence of the dynamic viscosity is represented by the power law
\begin{equation}
    \frac{\mu(T)}{\mu_\infty}
    =
    \left(\frac{T}{T_\infty}\right)^\omega,
    \label{eq_mu_power}
\end{equation}
where $\mu_\infty$ and $T_\infty$ are the free-stream viscosity and temperature, respectively, and $\omega$ is the viscosity exponent. For the VHS model, the free-stream mean free path is defined as
\begin{equation}
    \lambda_{\infty} =
    \frac{2(5-2\omega)(7-2\omega)}{15\sqrt{\pi}}
    \frac{\mu_{\infty}}
    {\rho_{\infty}\sqrt{2RT_\infty}},
\end{equation}
where $\rho_\infty$ is the free-stream density.
The 
calculations are performed with $\omega=0.72$. 

In the present study, the wedges serve only to generate the incident shocks. The region of interest is the local neighbourhood of RR, whose characteristic dimensions are much smaller than the macroscopic wedge length $w$. Consequently, a Knudsen number based on the characteristic size of the reflection region is substantially larger than the global value of $0.001$. This scale-based comparison should not be confused with a local gradient-length Knudsen number, which depends on the local mean free path and the gradients of the flow parameters. Nevertheless, the small global Knudsen number does not necessarily imply near-equilibrium conditions within the reflection region, where strong spatial gradients develop across the interacting shock-wave fronts.

Specular molecular reflection is imposed on the wedge surfaces, preventing the formation of a conventional viscous boundary layer. This boundary condition allows the non-equilibrium effects associated directly with the interaction of the incident and reflected shock-wave fronts to be examined without the additional influence of wedge boundary layers.

For a given molecular collision model, gas composition and boundary conditions, the local flow structure near RR is primarily governed by the incident-shock strength and the interaction geometry. In the present parameterization, these factors are characterized by $M_n$ and $\alpha$.

The computational domain is chosen so that the region of interest is sufficiently far from the trailing edges of the wedges and the external boundaries. This arrangement is intended to minimize the influence of expansion fans originating at the wedge trailing edges on the local flow structure near RR.

A Maxwellian distribution corresponding to the free-stream conditions is prescribed at the inflow boundary. At the outflow boundary, a free-outflow condition is applied: particles are allowed to leave the computational domain, with no particles entering through this boundary.

\section{DSMC implementation} 
\label{sec_methods}
The direct simulation Monte Carlo (DSMC) method \cite{Bird_book} treats the gas flow as an ensemble of simulated particles. Each simulated particle represents a large number ($\sim 10^{12}-10^{20}$) of real gas molecules. At each time step $\Delta t$, the simulation is divided into two independent stages: free molecular motion and collisional relaxation. During the first stage, particles move over distances proportional to their velocities. If a particle encounters a solid surface, its reflection is modelled according to the prescribed gas--surface interaction law. During the second stage, molecular collisions are simulated stochastically within each computational cell, neglecting the relative positions of the particles. Spatial distributions of macroscopic flow parameters, such as velocity, density and temperature, are obtained by averaging sampled molecular properties in each cell over a time interval after a steady state has been reached.

DSMC can be regarded as a Monte Carlo method for numerically solving the Boltzmann equation in the limit as the number of simulated particles tends to infinity (see \citep{IvanovRogasinsky}). DSMC solutions for strong shock fronts are known to agree with direct numerical solutions of the Boltzmann equation (see, for example, \citep{ShockWaves_2015}).

The DSMC calculations were performed using the Statistical Modeling In Low-density Environment (SMILE++) software \citep{KashkovskySMILEpp, IvanovSMILE2010}, which employs the majorant frequency scheme \citep{IvanovRogasinsky}. The variable hard sphere (VHS) model was used for elastic molecular collisions. This model is consistent with a power-law temperature dependence of viscosity with exponent $\omega$.

At the start of each steady RR simulation, the domain is populated with simulated particles according to the Maxwellian velocity distribution corresponding to the free-stream conditions. The computation then proceeds until a steady state is reached, after which molecular properties are sampled to obtain the macroscopic flow fields.

\FloatBarrier
\section{Results and discussion} \label{sec_results}
The following nondimensionalization of the flow parameters is used in the section:
\begin{align}
  \hat v_i &= \frac{v_i}{C_\infty}, &
  \hat x &= \frac{x}{\lambda_\infty}, &
  \hat y &= \frac{y}{\lambda_\infty}, &
  \hat T &= \frac{T}{T_\infty}, \nonumber\\
  \hat\rho &= \frac{\rho}{\rho_\infty}, &
  \hat p &= \frac{p}{\rho_\infty C_\infty^2}
          = \frac12 \hat\rho \hat T, &
  \hat q_i &= \frac{q_i}{\rho_\infty C_\infty^3}, &
  \hat\sigma_{ij} &= \frac{\sigma_{ij}}
  {\rho_\infty C_\infty^2}.
\label{eq_Nondimensionalization}
\end{align}
where $C_\infty=\sqrt{2RT_\infty}$ is the absolute value of the free-stream most probable peculiar molecular velocity. The $\infty$ subscript denotes the free-stream values. Only nondimensional variables \eqref{eq_Nondimensionalization} are used below in the present section with the "hat" symbol omitted except for the nondimensional coordinates in figures where they are denoted as $x / \lambda_{\infty}$ and $y / \lambda_{\infty}$. 

\subsection{Two-dimensional flow near the regular-reflection point}
\label{subsec_2D_results}

\subsubsection{Overall flow-field structure}
\label{subsubsec_2D_fields}

\begin{figure}[!htbp]
\centering
\RRthreepanels
  {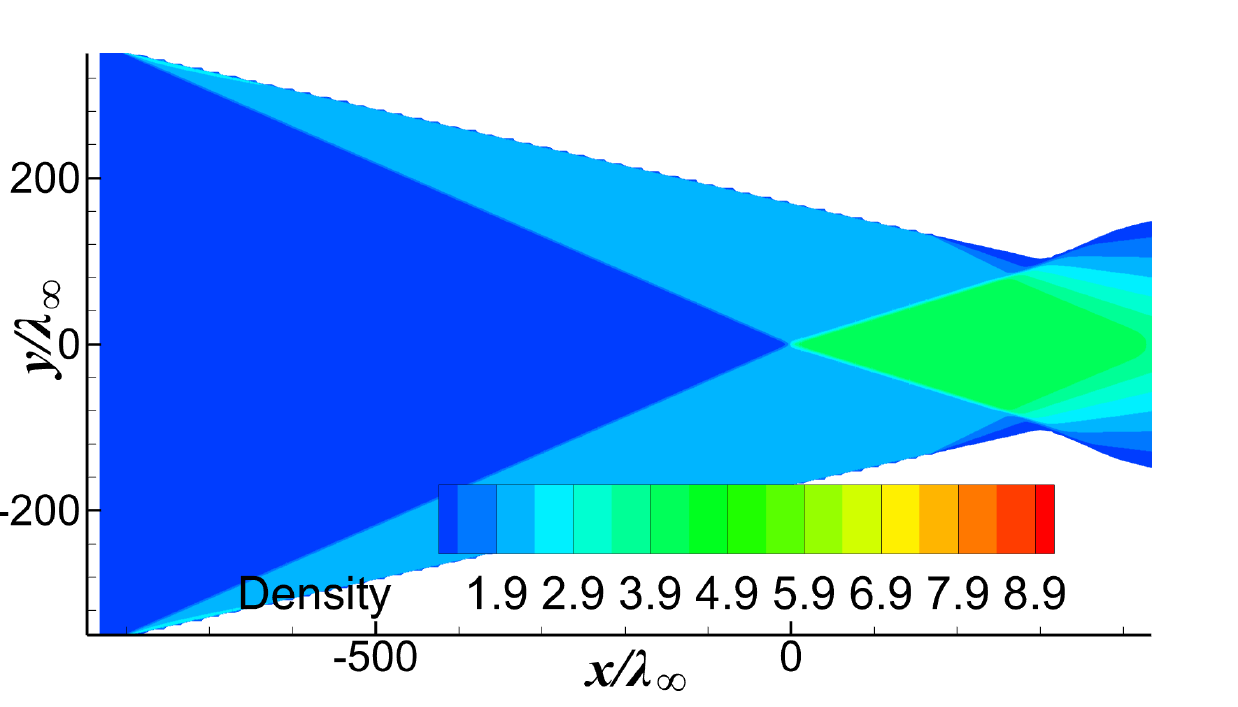}
  {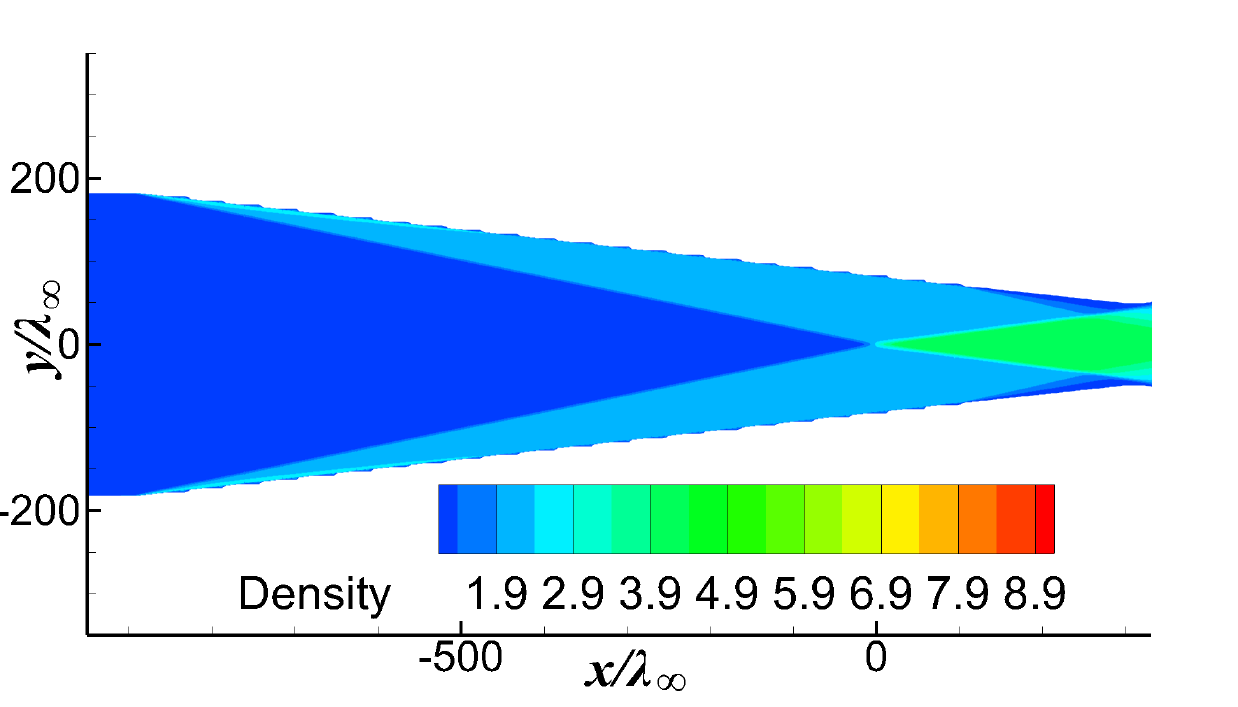}
  {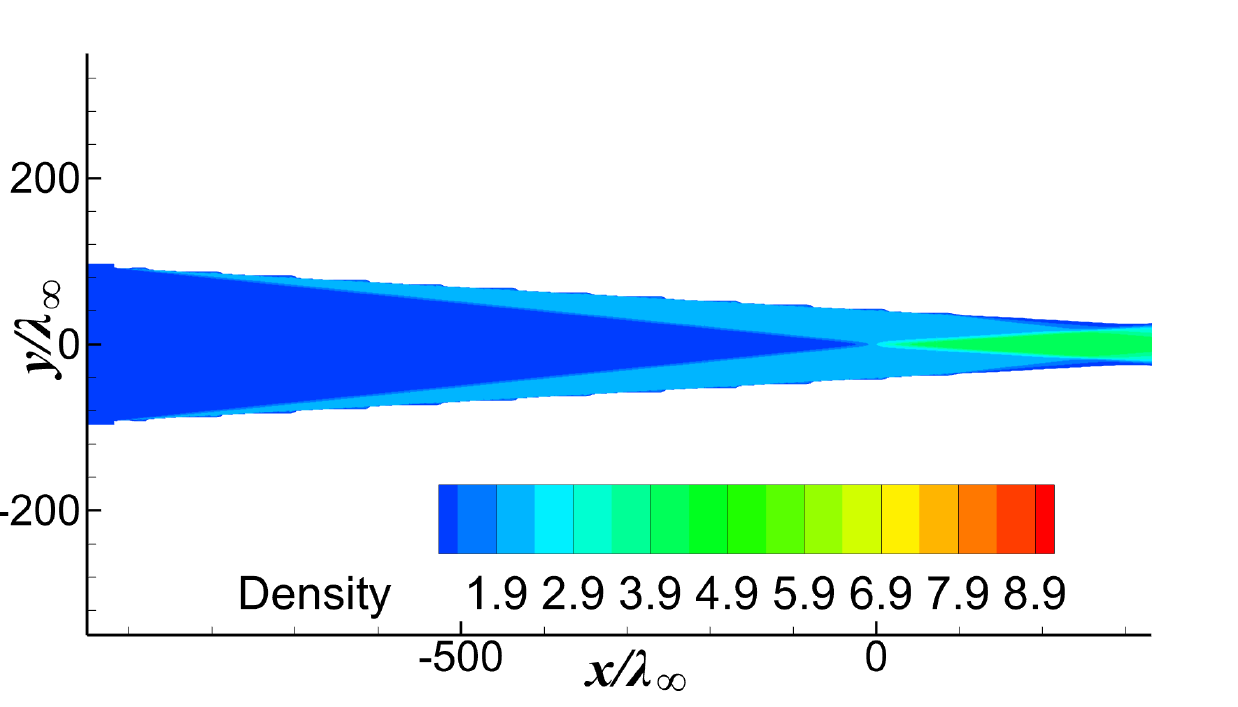}
\caption{Density fields for $M_n=2.0$ series.
$M_\infty=5$ (a), $M_\infty=10$ (b) and
$M_\infty=20$ (c).}
\label{fig_Rho_full_Mn2}
\end{figure}

\begin{figure}[!htbp]
\centering
\RRthreepanels
  {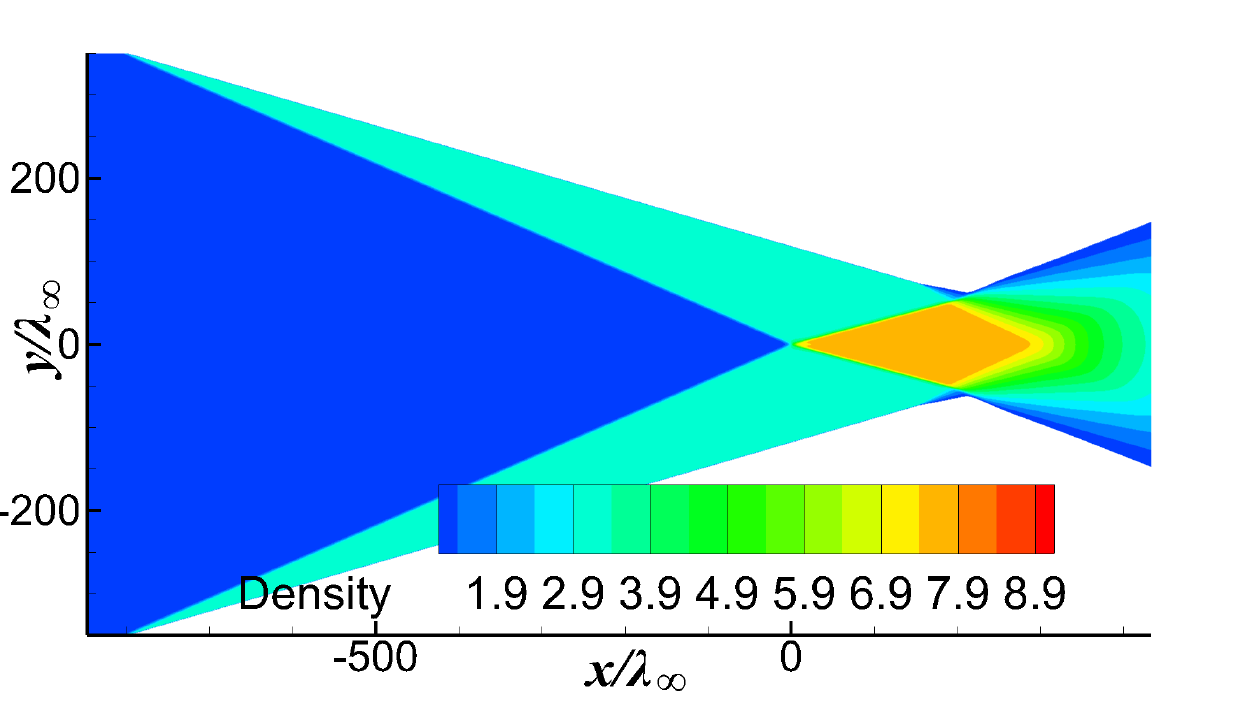}
  {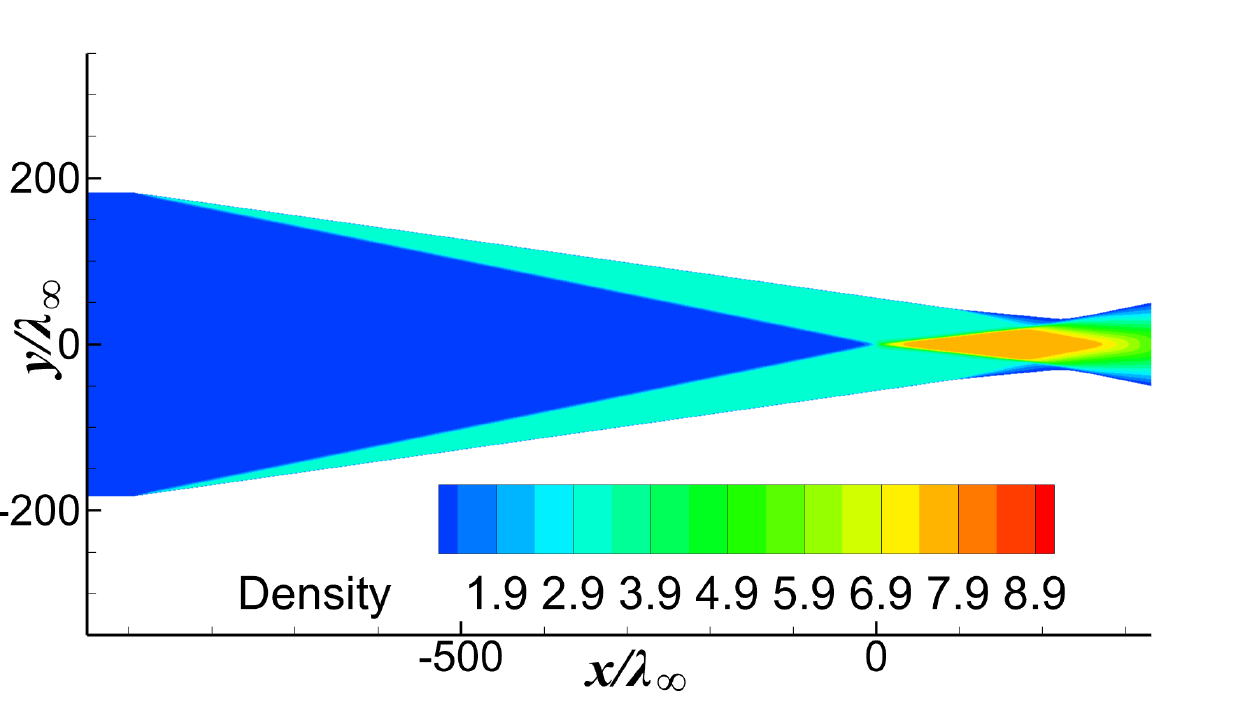}
  {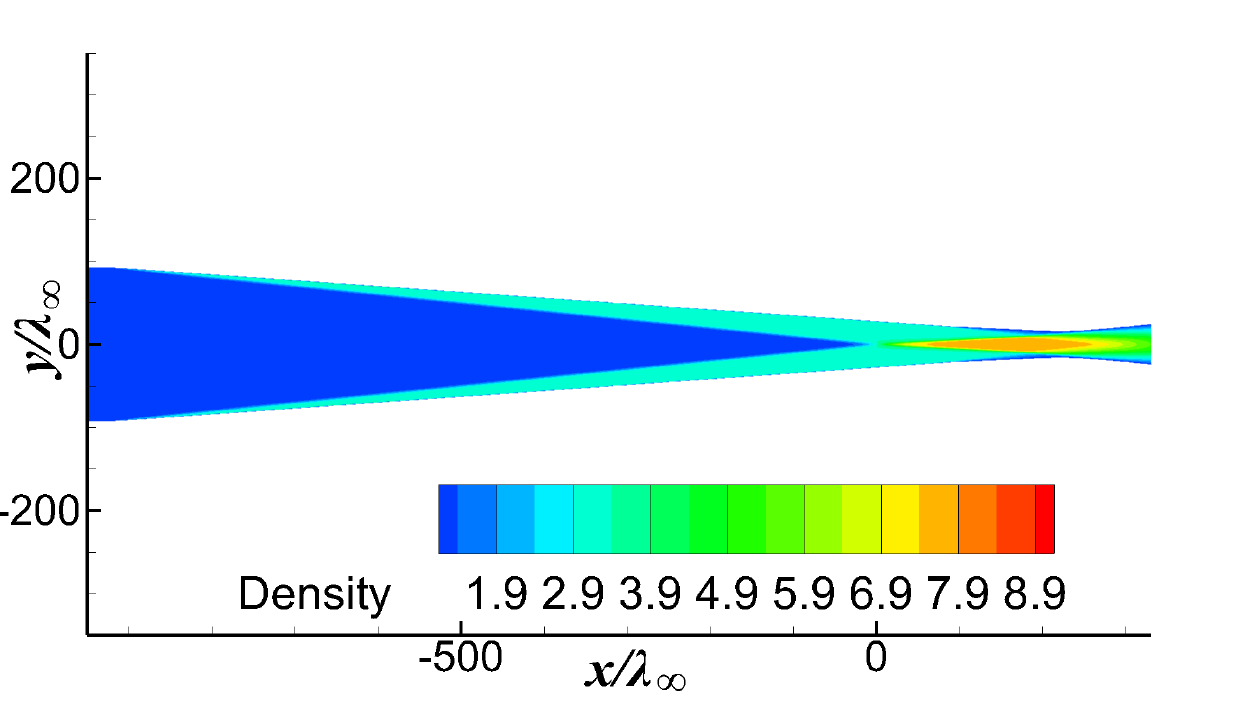}
\caption{Density fields for $M_n=4.0$ series.
$M_\infty=10$ (a), $M_\infty=20$ (b) and
$M_\infty=40$ (c).}
\label{fig_Rho_full_Mn4}
\end{figure}

\begin{figure}[!htbp]
\centering
\RRthreepanels
  {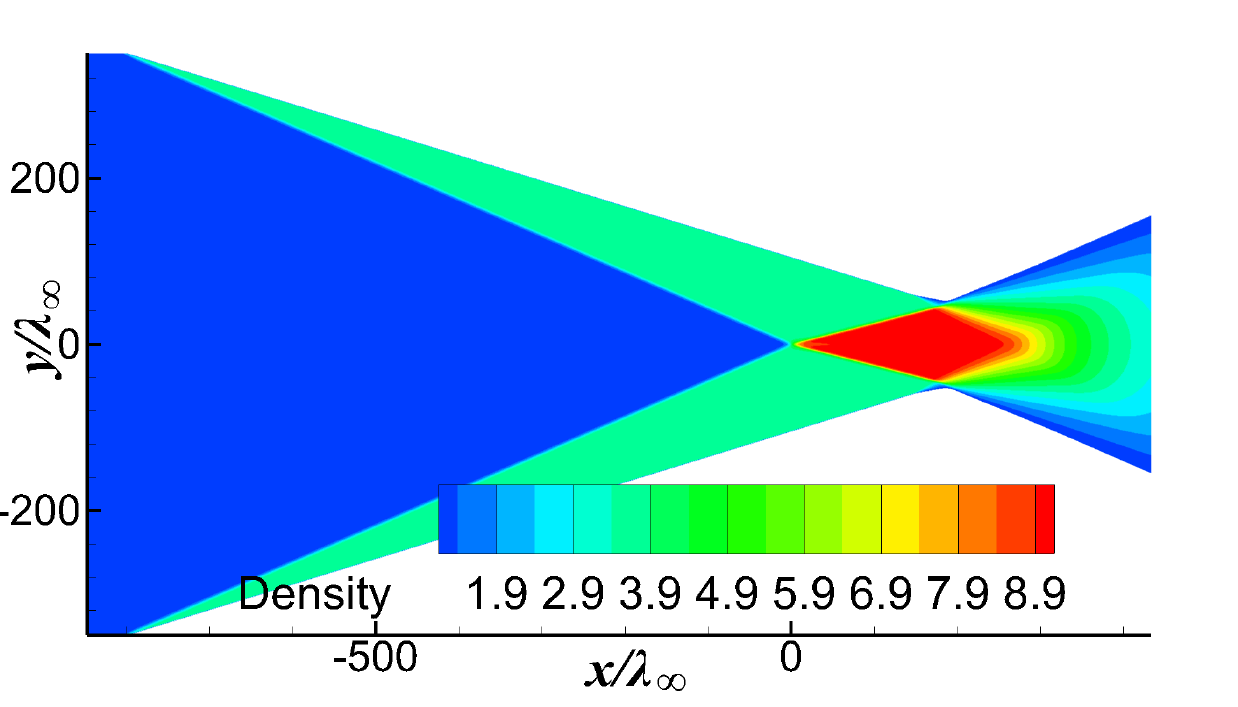}
  {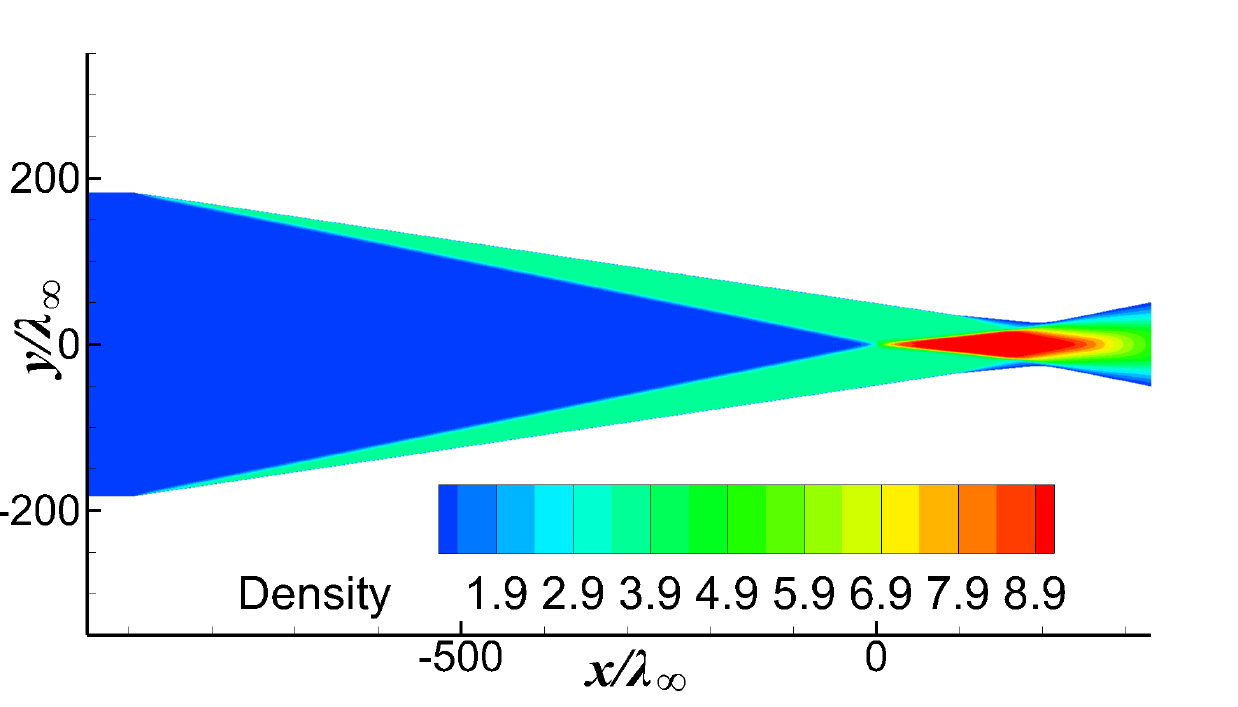}
  {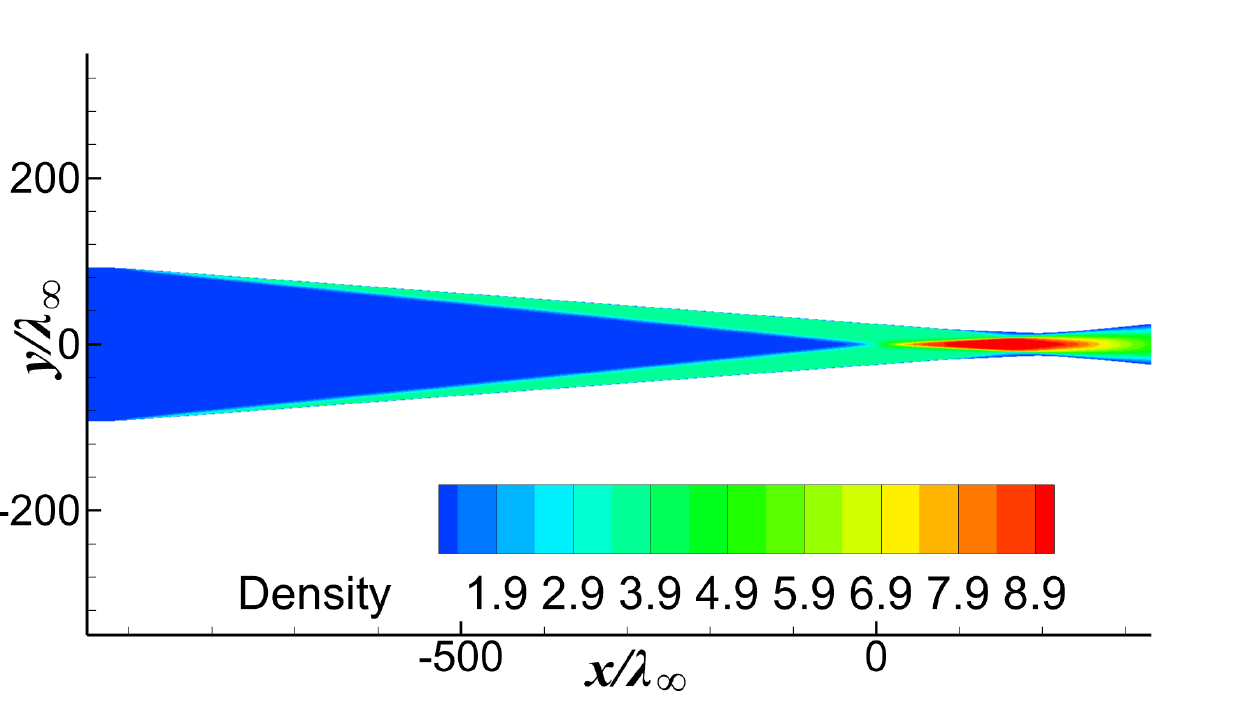}
\caption{Density fields for $M_n=8.0$ series.
$M_\infty=20$ (a), $M_\infty=40$ (b) and
$M_\infty=80$ (c).}
\label{fig_Rho_full_Mn8}
\end{figure}

Figures~\ref{fig_Rho_full_Mn2}--\ref{fig_Rho_full_Mn8} show the two-dimensional density fields for all nine cases listed in table~\ref{tab_cases}. Figures~\ref{fig_Rho_full_Mn2}, \ref{fig_Rho_full_Mn4} and \ref{fig_Rho_full_Mn8} correspond to the $M_n=2$, $4$ and $8$ series, respectively. Within each series, panels (a)--(c) show decreasing incidence angles $\alpha$, from $23.578^\circ$ to $5.739^\circ$. 
A common colour scale is used for all nine density fields to facilitate direct comparison of the compression levels.

The coordinate origin $(0,0)$ is the reflection point in the corresponding inviscid solution. In the kinetic solution, the incident and reflected shock-wave fronts have finite thicknesses, so no unique geometric intersection point exists. Hereafter, $(0,0)$ is used as the nominal centre of the interaction region.

The characteristic two-shock structure of RR is retained in all nine cases: the incident and reflected shock-wave fronts interfere near the plane of symmetry without forming a Mach stem. Near the nominal reflection point, the regions influenced by the IS and RS overlap and form a single two-dimensional transition region.

At fixed $M_n$, an increase in $M_\infty$ corresponds to a decrease in $\alpha$. The incident and reflected shock waves become more nearly aligned with the plane of symmetry. At the same time, their interaction region becomes longer in the streamwise direction and narrower in the transverse direction. At the smallest angles considered, a narrow and extended transition region forms along the plane of symmetry.

This trend occurs in all three series. In particular, the $M_n=2$ results demonstrate that the elongation of the interaction region as $\alpha$ decreases persists at a substantially lower normal incident-shock strength than in the $M_n=4$ and $8$ series.

At fixed $M_n$, the thermodynamic jumps across the incident shock remain unchanged. In particular,
\begin{equation}
    \frac{\rho_2}{\rho_1}
    =
    \frac{(\gamma+1)M_n^2}
    {(\gamma-1)M_n^2+2}.
    \label{eq_rho12_Mn}
\end{equation}
For the three series, the density ratios are $2.286$, $3.368$ and $3.821$ at $M_n=2$, $4$ and $8$, respectively. Hence, the changes in the two-dimensional structure within a series are primarily associated with the interaction geometry and the conditions ahead of the reflected shock, rather than with changes in the strength of the incident shock itself.

The equilibrium state 3 also depends only weakly on $\alpha$ within each series. According to the corresponding inviscid solutions, $\rho_3/\rho_1$ ranges from $4.26$ to $4.34$, from $7.62$ to $7.81$, and from $9.12$ to $9.36$ for $M_n=2$, $4$ and $8$, respectively. The equilibrium density varies by less than $3\%$ within each series, whereas the geometry of the transition region changes much more substantially.

Comparing different series at the same incidence angle instead reveals the effect of normal shock strength. As $M_n$ increases from $2$ to $4$ and $8$, the density jumps across both the incident and reflected shocks increase, and the equilibrium density behind the shock system rises accordingly.

\subsubsection{The vicinity of the reflection point}
\label{subsubsec_reflection_region}

\begin{figure}[!htbp]
\centering
\RRverticalpanels
  {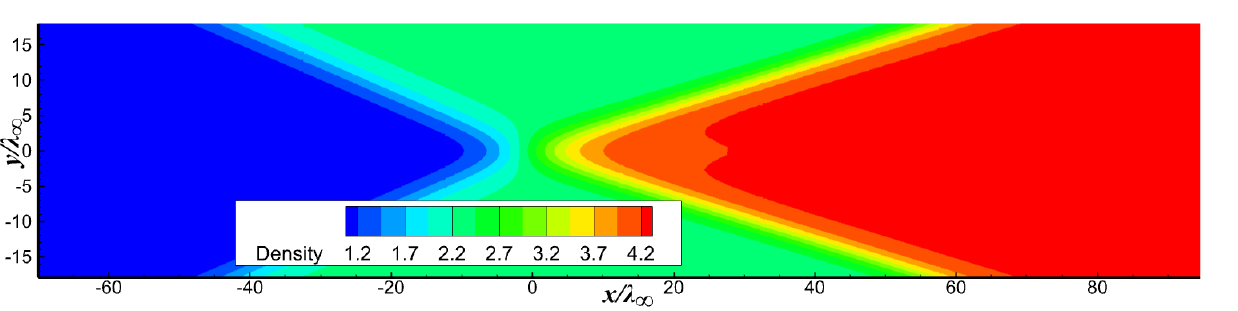}
  {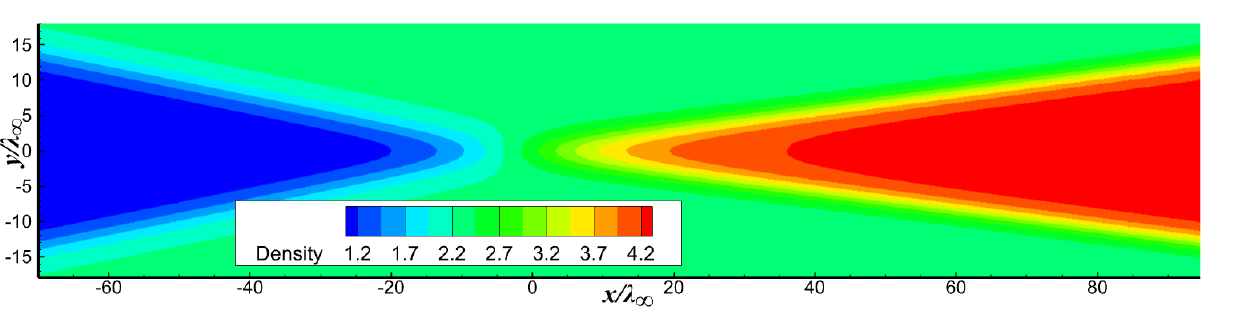}
  {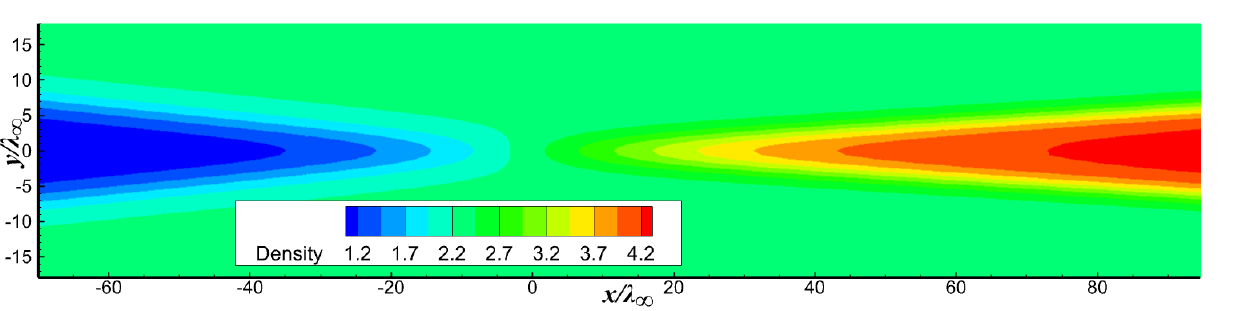}
\caption{Density fields for $M_n=2.0$.
$M_\infty=5$ (a), $M_\infty=10$ (b)
and $M_\infty=20$ (c).}
\label{fig_Rho_Mn2}
\end{figure}

\begin{figure}[!htbp]
\centering
\RRverticalpanels
  {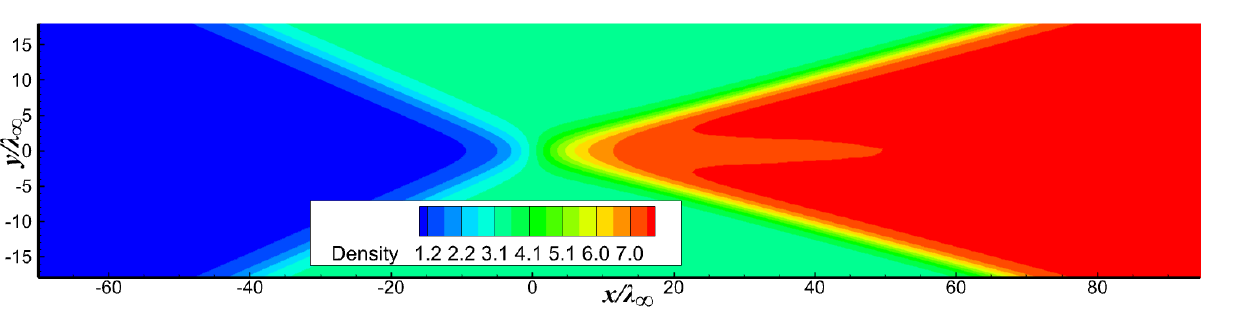}
  {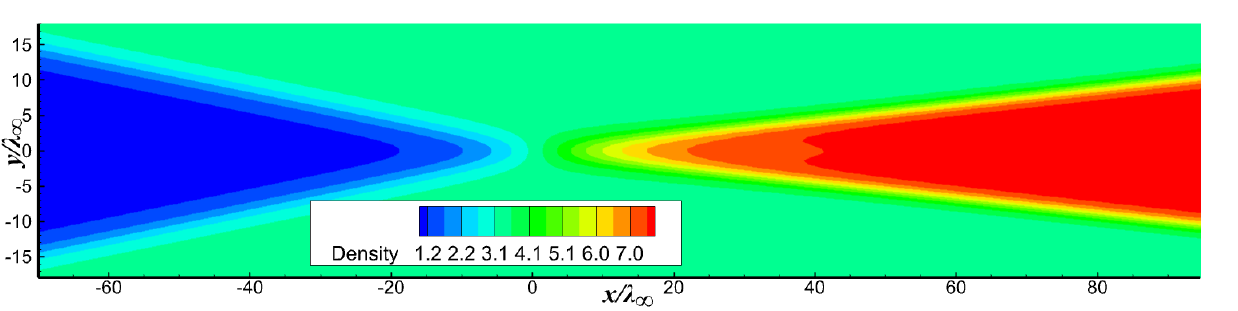}
  {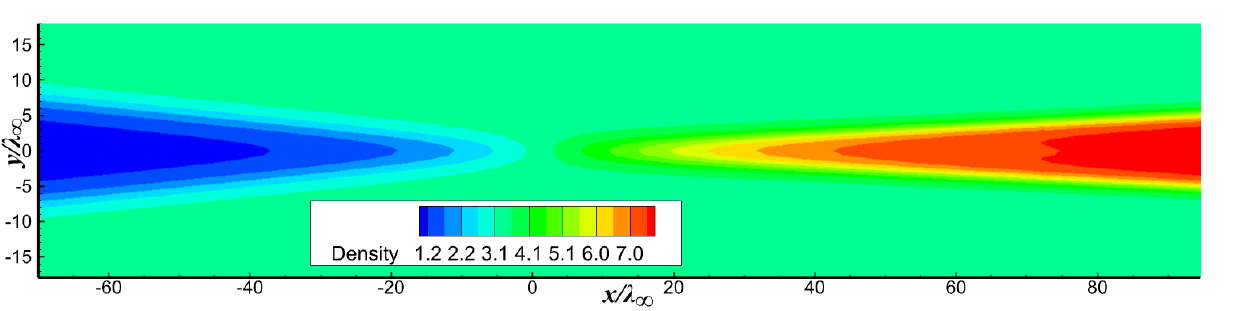}
\caption{Density fields for $M_n=4.0$.
$M_\infty=10$ (a), $M_\infty=20$ (b)
and $M_\infty=40$ (c).}
\label{fig_Rho_Mn4}
\end{figure}

\begin{figure}[!htbp]
\centering
\RRverticalpanels
  {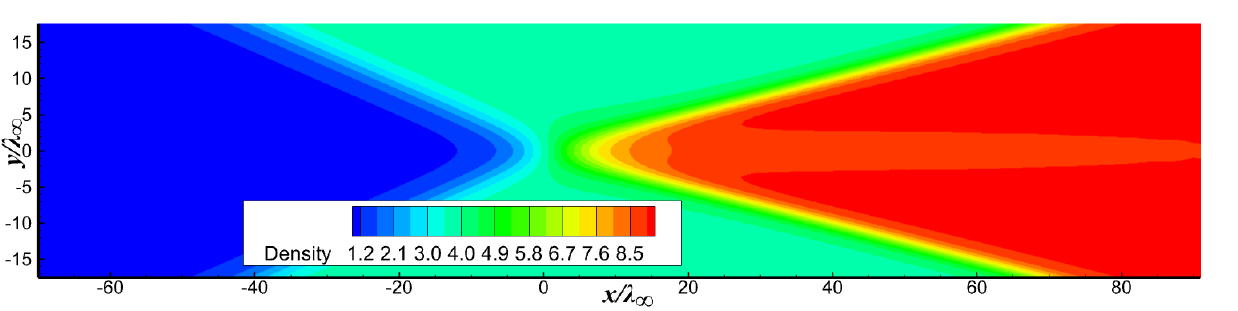}
  {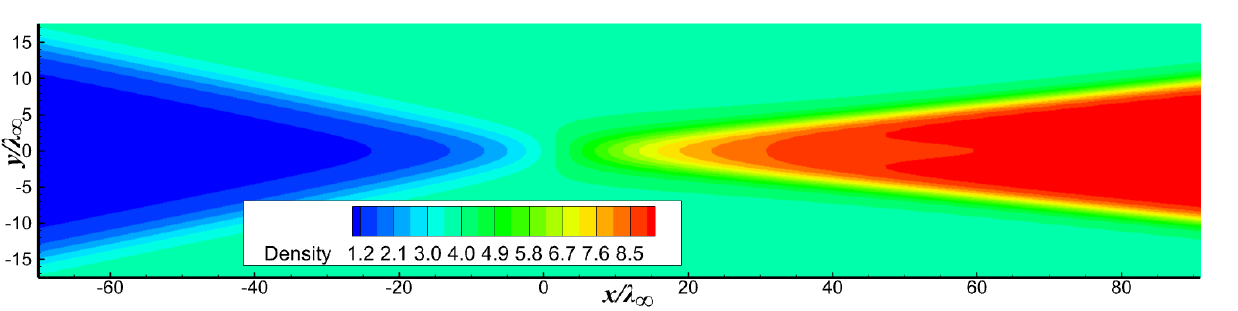}
  {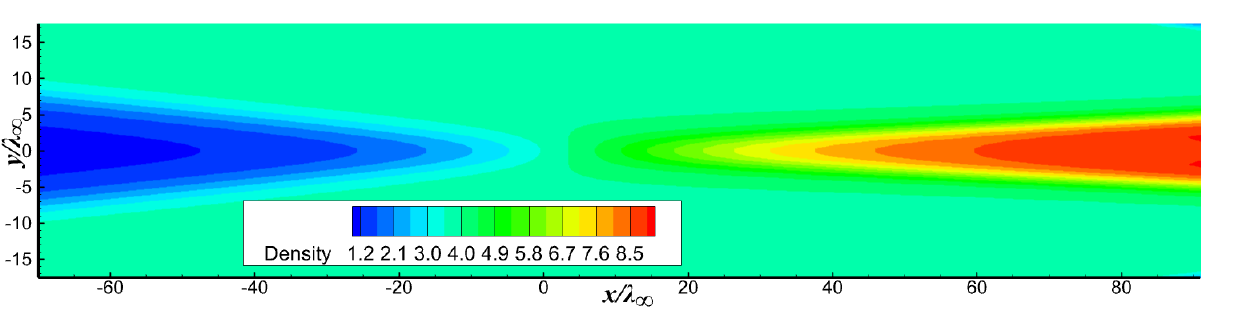}
\caption{Density fields for $M_n=8.0$.
$M_\infty=20$ (a), $M_\infty=40$ (b)
and $M_\infty=80$ (c).}
\label{fig_Rho_Mn8}
\end{figure}

For a more detailed analysis of the interaction region, figures~\ref{fig_Rho_Mn2}--\ref{fig_Rho_Mn8} show zoomed views of the density fields near RR. In every case, the coordinates are normalized by the free-stream mean free path $\lambda_\infty$. 
To resolve the internal structure of the shock-wave fronts, the colour scales of the enlarged density fields are adjusted separately for each $M_n$ series, while remaining identical for the three incidence angles within a series. The same convention is used for the temperature fields shown below. Consequently, identical colours in different $M_n$ series do not necessarily correspond to the same density or temperature values.

\FloatBarrier

\FloatBarrier

The enlarged density fields illustrate the finite thicknesses of the incident and reflected shock-wave fronts and the way in which they overlap. Near the nominal reflection point, the two shock fronts can no longer be regarded as independent; instead, they form a common region where the flow changes continuously from the state ahead of the incident shocks to the state behind the reflected shocks.

The main geometric trend identified in the full flow fields is particularly evident at this enlarged scale. As $\alpha$ decreases, the region of mutual influence of the IS and RS becomes substantially longer along the $x$ axis, while its transverse extent decreases. The same behaviour occurs for all three values of $M_n$. Thus, the narrow, extended transition region formed at small incidence angles is not limited to cases with high $M_n$, but also occurs for the $M_n=2$ series.

Increasing $M_n$ also produces a larger range of density variation. Far downstream of the reflected shocks, the density approaches the equilibrium value for state 3, which rises from $\rho_3/\rho_1\simeq4.3$ for $M_n=2$ to approximately $7.7$ for $M_n=4$ and $9.2$ for $M_n=8$. Within each individual series, the variation of these equilibrium values with $\alpha$ remains small compared with the changes in the geometry of the transition region.

\begin{figure}[!htbp]
\centering
\RRverticalpanels
  {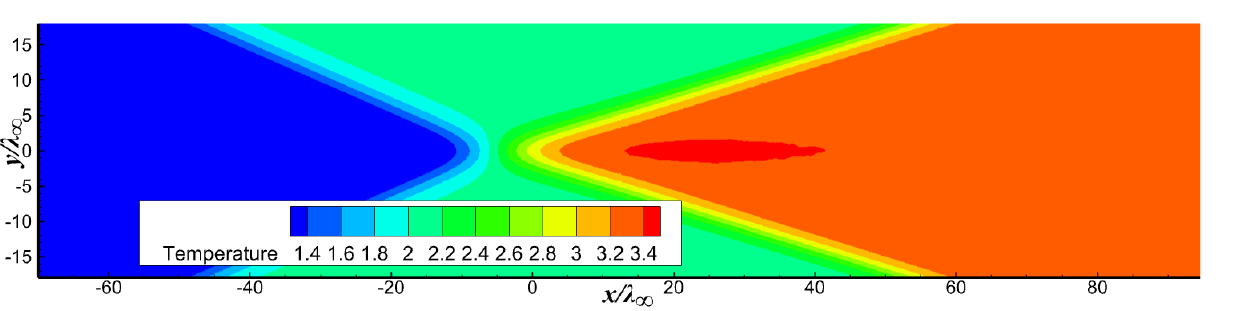}
  {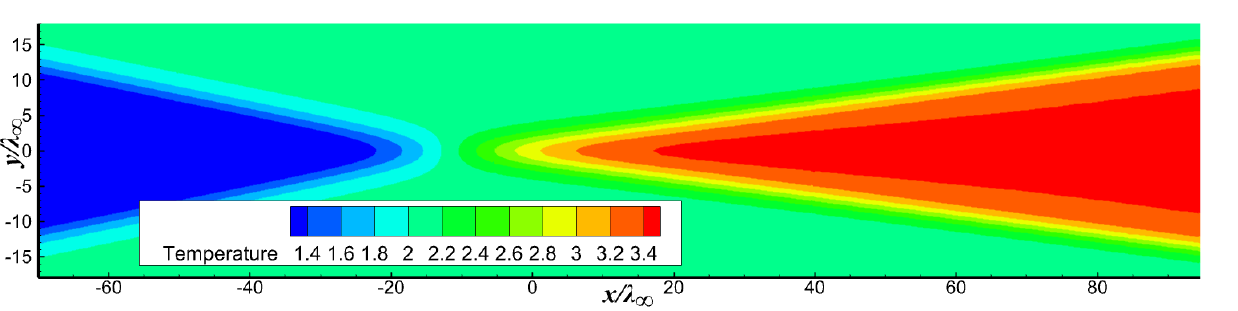}
  {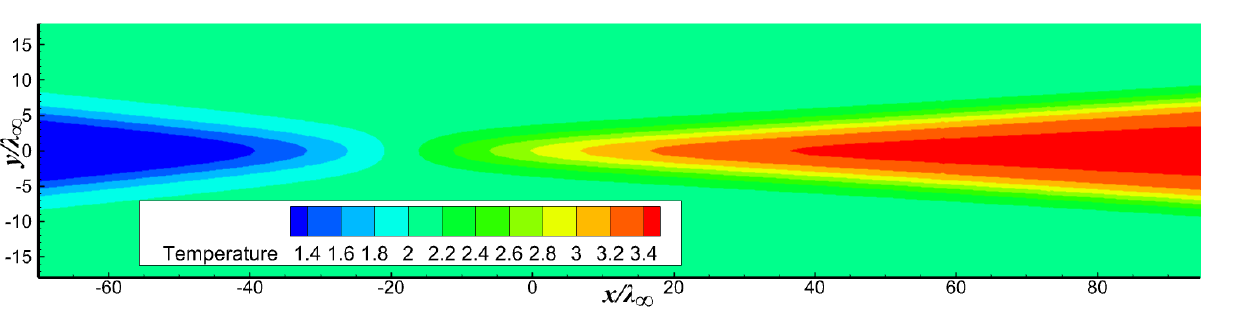}
\caption{Temperature fields for $M_n=2.0$.
$M_\infty=5$ (a), $M_\infty=10$ (b)
and $M_\infty=20$ (c).}
\label{fig_T_Mn2}
\end{figure}

\begin{figure}[!htbp]
\centering
\RRverticalpanels
  {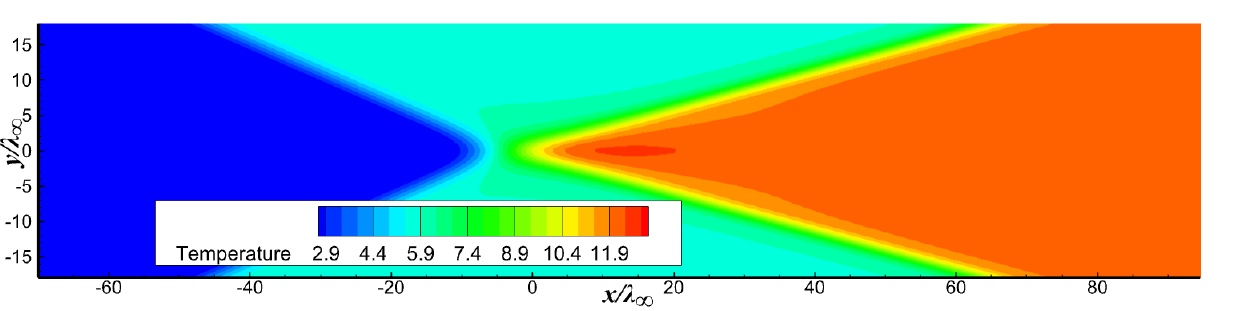}
  {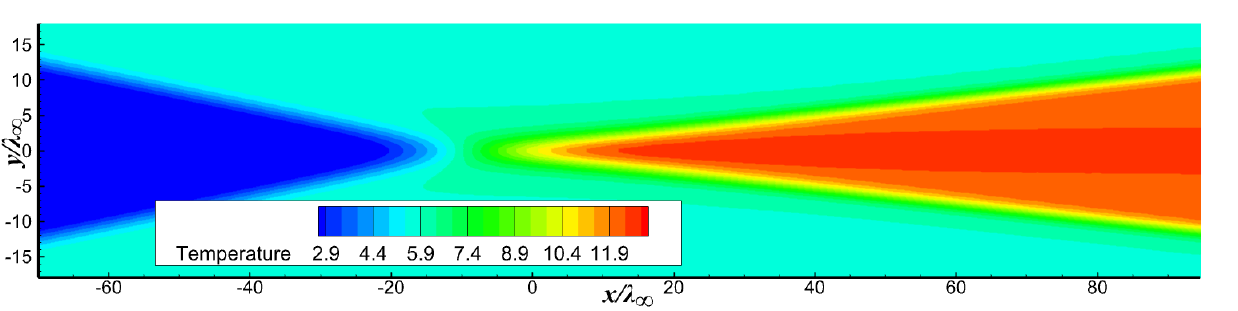}
  {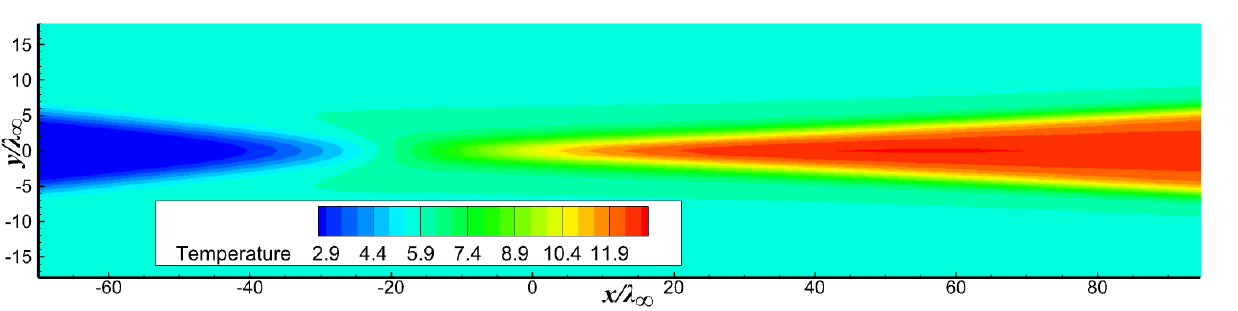}
\caption{Temperature fields for $M_n=4.0$.
$M_\infty=10$ (a), $M_\infty=20$ (b)
and $M_\infty=40$ (c).}
\label{fig_T_Mn4}
\end{figure}

\begin{figure}[!htbp]
\centering
\RRverticalpanels
  {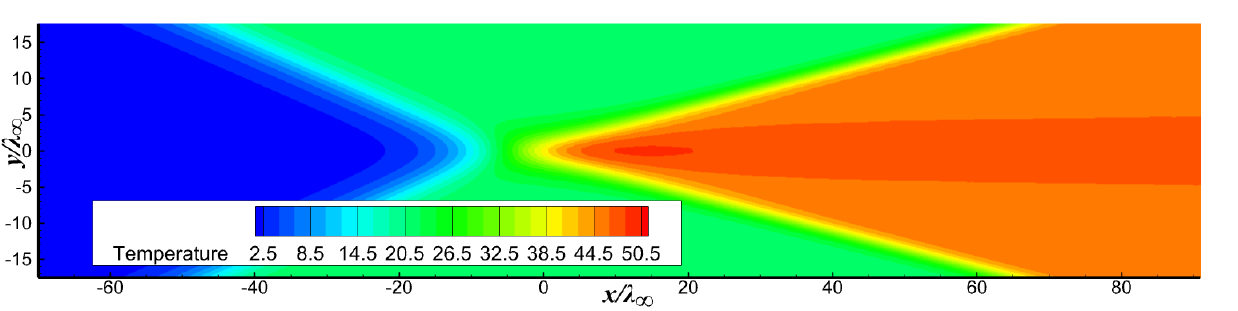}
  {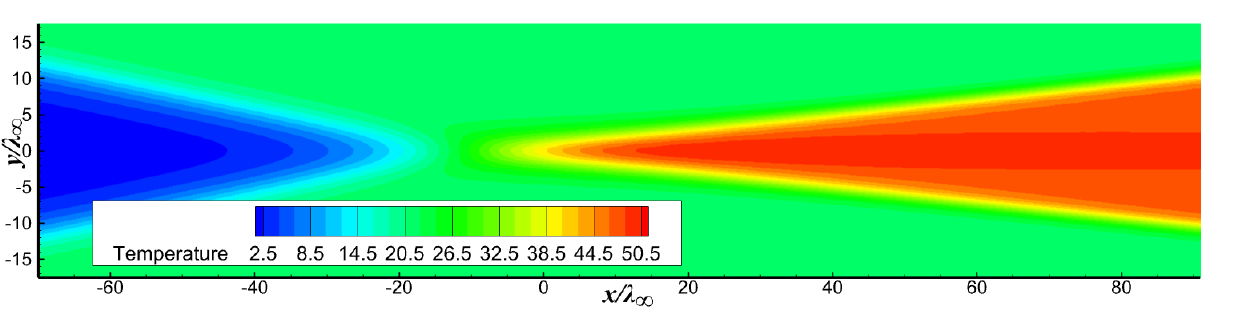}
  {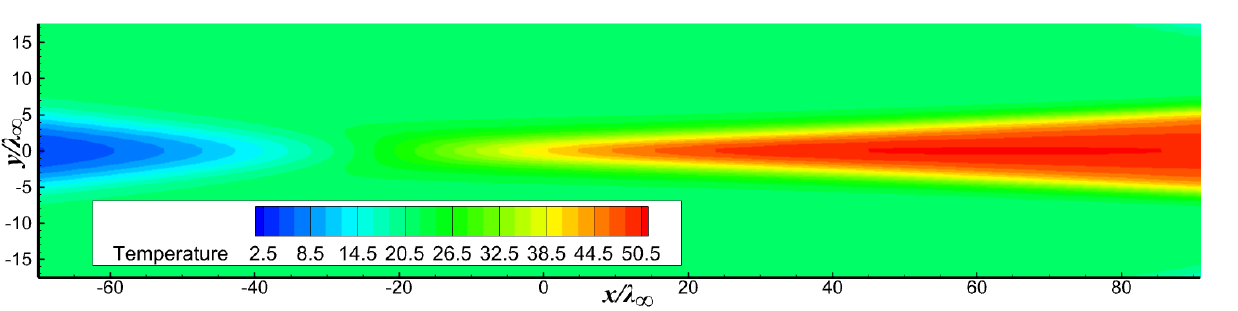}
\caption{Temperature fields for $M_n=8.0$.
$M_\infty=20$ (a), $M_\infty=40$ (b)
and $M_\infty=80$ (c).}
\label{fig_T_Mn8}
\end{figure}

Another characteristic feature of the RR region appears in the total translational-temperature field. The corresponding distributions for the three series are shown in figures~\ref{fig_T_Mn2}--\ref{fig_T_Mn8}. As for density, decreasing $\alpha$ stretches the temperature-transition region along the plane of symmetry while narrowing it in the transverse direction. However, changes in $M_n$ have a much stronger effect on the absolute temperature levels than changes in $\alpha$ within a single series.

\FloatBarrier

For the cases considered, the temperature ratio $T_3/T_1$ ranges from $3.37$ to $3.43$, from $12.00$ to $12.45$, and from $46.30$ to $48.34$ for $M_n=2$, $4$ and $8$, respectively. Thus, as with density, the equilibrium temperature varies relatively little within a series, while the spatial structure of the transition region changes markedly as $\alpha$ decreases.

At the largest incidence angle, $\alpha=23.578^\circ$, a localized high-temperature region is observed close to the plane of symmetry downstream of the overlap of the shock-wave fronts. This effect is relatively weak for $M_n=2$, but the contrast with the surrounding flow becomes much more pronounced for $M_n=4$ and especially $M_n=8$. This behaviour is consistent with the previously reported non-Rankine--Hugoniot zone \cite{BondarShoevTimokhin2024}, in which the local gas state differs from equilibrium state 3.

As $\alpha$ decreases, the two-dimensional temperature overshoot becomes less localized: the high-temperature region extends along the plane of symmetry and develops into a long temperature wake. This evolution occurs in all three series and follows the general geometric rearrangement established from the density fields.

The two-dimensional density and total translational-temperature fields therefore reveal two robust parametric trends. Decreasing $\alpha$ at fixed $M_n$ primarily extends the transition region in the streamwise direction and narrows it transversely, whereas increasing $M_n$ at comparable incidence angles substantially increases the levels of compression and heating.

\subsection{The plane of symmetry}
\label{subsec_symmetry_plane}

The two-dimensional fields demonstrate that varying the parameters substantially rearranges the geometry of the RR region. To compare the cases quantitatively, we now examine the distributions of macroscopic flow parameters along the plane of symmetry, $y=0$.

\subsubsection{Normalized density, temperature and streamwise-velocity distributions}
\label{subsubsec_symmetry_plane_density}

The total changes in density, temperature and streamwise velocity between states 1 and 3 differ substantially among the $M_n$ series. To compare the shapes of the transition profiles independently of these equilibrium-state differences, the macroscopic parameters are normalized by their complete changes between states 1 and 3. For density and temperature, we introduce
\begin{equation}
    \rho_N
    =
    \frac{\rho-\rho_1}{\rho_3-\rho_1},
    \qquad
    T_N
    =
    \frac{T-T_1}{T_3-T_1},
    \label{eq_rho_T_normalized}
\end{equation}
Thus, $\rho_N=T_N=0$ in state 1 and $\rho_N=T_N=1$ in equilibrium state 3. For streamwise velocity, we use
\begin{equation}
    v_{x,N}
    =
    \frac{v_x-v_{x3}}
    {v_{x1}-v_{x3}},
    \label{eq_vx_normalized}
\end{equation}
This normalization preserves the direction of change of the original velocity profile: $v_{x,N}=1$ in state 1 and $v_{x,N}=0$ in state 3. Negative $v_{x,N}$ therefore indicates a local streamwise velocity below its equilibrium downstream value $v_{x3}$.

\begin{figure}[!htbp]
\centering
\RRprofileverticalpanels
  {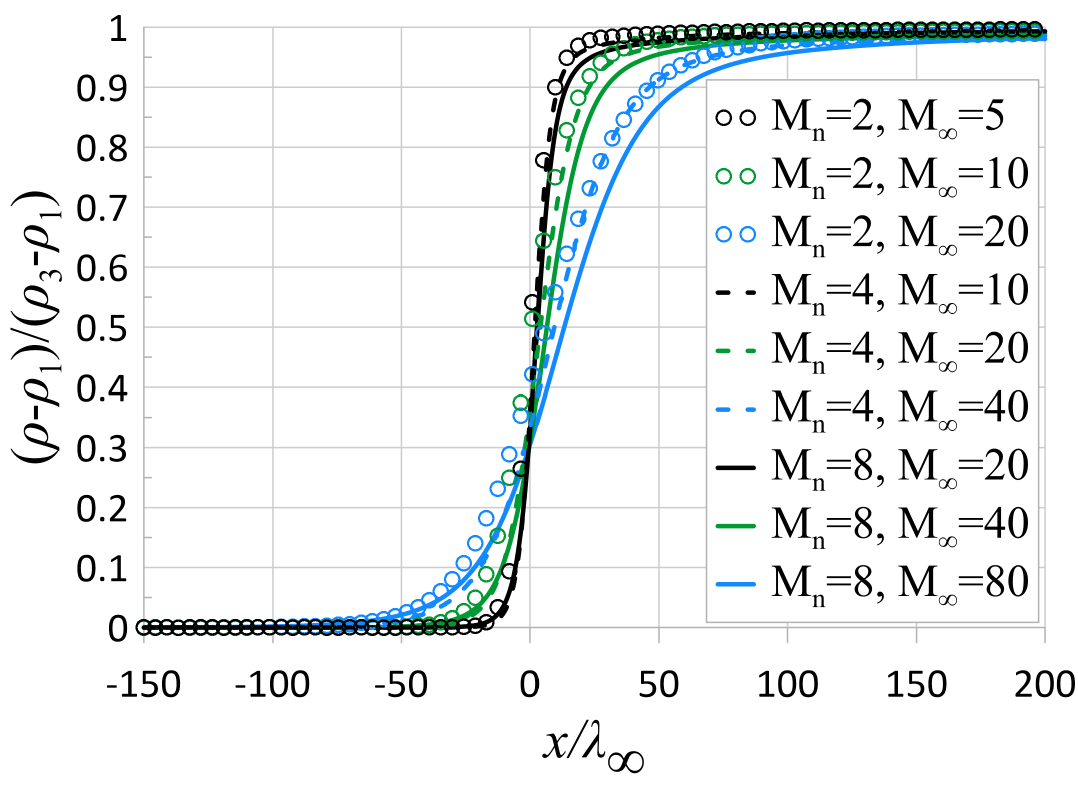}
  {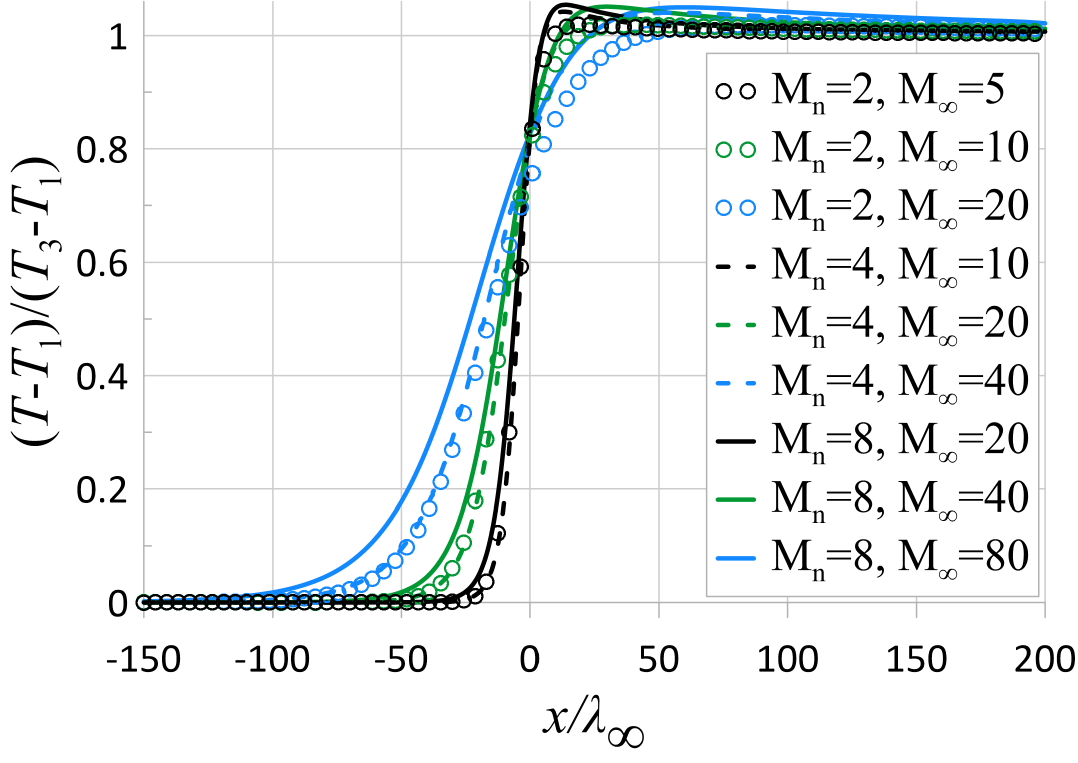}
  {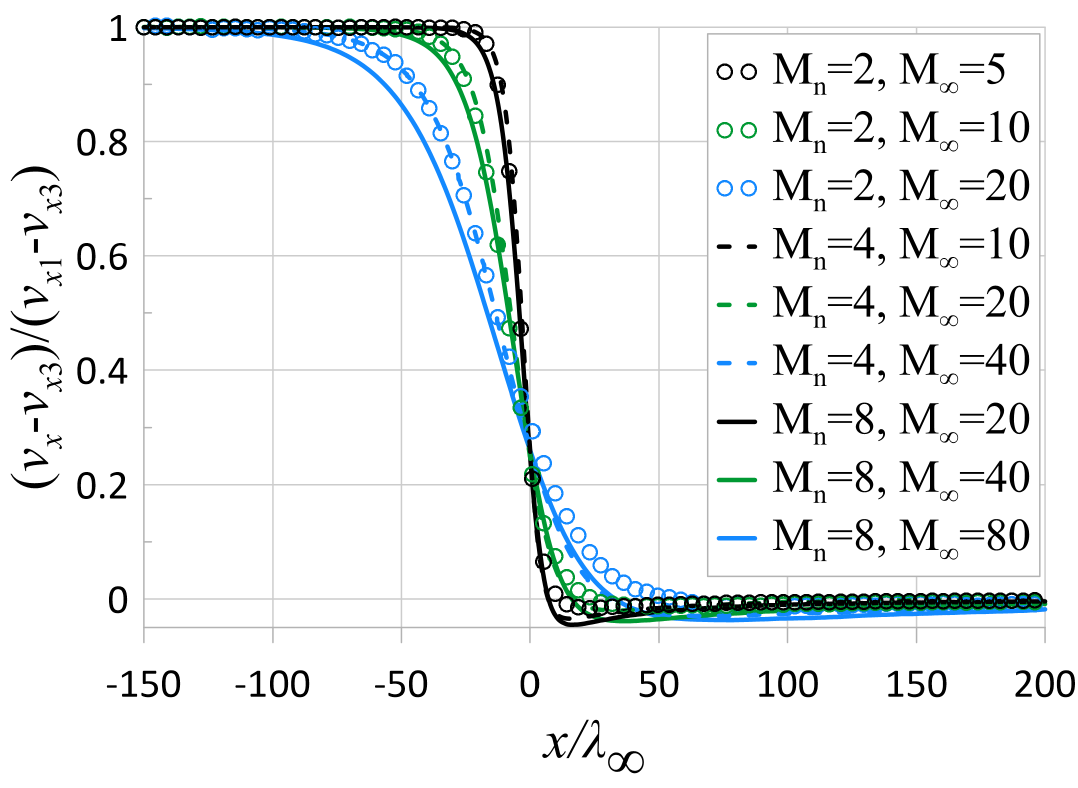}
\caption{Normalized density (a), temperature (b),
and streamwise velocity (c) distributions along
the symmetry plane for all the considered cases.}
\label{fig_shock}
\end{figure}

Figure~\ref{fig_shock} shows normalized density, temperature and streamwise-velocity profiles along the plane of symmetry for all nine cases. Normalization by the total changes between states 1 and 3 allows the shapes of the transition profiles to be compared directly, despite large differences in shock strength. All three quantities show a clear grouping of profiles by incidence angle $\alpha$. At a given $\alpha$, the profiles for $M_n=2$, $4$ and $8$ are relatively close, whereas decreasing $\alpha$ produces a marked increase in the width of the transition region.

At the largest angle, $\alpha=23.578^\circ$, the variations of the macroscopic parameters are concentrated within a relatively narrow region around the nominal reflection point. At $\alpha=11.537^\circ$, the transition becomes more extended along the plane of symmetry, and at $\alpha=5.739^\circ$ the broadest profiles occur, with long upstream and downstream portions. This trend is consistent with the two-dimensional density and temperature fields, which show that the interaction region extends along the $x$ axis as $\alpha$ decreases.

This behaviour is particularly clear in the normalized density profiles in figure~\ref{fig_shock}(a). After accounting for the different overall density jumps, $M_n$ has a relatively weak effect on the normalized profile shape, whereas decreasing $\alpha$ at fixed $M_n$ causes a substantially greater increase in its extent along the symmetry plane.

To quantify this extension of the density profile, we define a density-transition width along the plane of symmetry. Let $x_{0.1}$ and $x_{0.9}$ be the dimensional positions where the normalized density reaches $\rho_N=0.1$ and $\rho_N=0.9$, respectively. The dimensionless transition width is then
\begin{equation}
    L_\rho = \hat x_{0.9}-\hat x_{0.1} = \frac{x_{0.9}-x_{0.1}}{\lambda_\infty}.
\label{eq_Lrho}
\end{equation}
Thus, $L_\rho$ characterizes the extent of the density transition along the plane of symmetry in units of the free-stream mean free path.

\begin{table}[ht]
\centering
\caption{Density-transition width along the plane of symmetry for $\omega=0.72$.}
\label{tab_Lrho_072}
\begin{tabular}{rrrrr}
\toprule
$M_n$ & $M_\infty$ & $\alpha$, $^\circ$ &
$L_\rho$ & $L_\rho\sin\alpha$ \\
\midrule
2 &  5 & 23.578 & 17.5 & 7.01 \\
2 & 10 & 11.537 & 36.5 & 7.31 \\
2 & 20 &  5.739 & 73.5 & 7.35 \\
\midrule
4 & 10 & 23.578 & 15.9 & 6.37 \\
4 & 20 & 11.537 & 33.6 & 6.72 \\
4 & 40 &  5.739 & 65.7 & 6.57 \\
\midrule
8 & 20 & 23.578 & 19.6 & 7.85 \\
8 & 40 & 11.537 & 40.3 & 8.05 \\
8 & 80 &  5.739 & 80.4 & 8.04 \\
\bottomrule
\end{tabular}
\end{table}

The values of $L_\rho$ are presented in table~\ref{tab_Lrho_072}.
Within each series, decreasing $\alpha$ produces an almost proportional increase in $L_\rho$. Since $M_n=M_\infty\sin\alpha$ remains constant, successive reductions of $\sin\alpha$ from $0.4$ to $0.2$ and $0.1$ are accompanied by an approximately twice increase in $L_\rho$ at each step.

The quantity $L_\rho\sin\alpha$ varies relatively little within each series. Therefore, over the parameter range considered, the results are consistent with the approximate scaling $L_\rho\propto1/\sin\alpha$ at fixed $M_n$ and $\omega$. The proportionality coefficient may depend on both parameters. Angular stretching of this kind is expected for an isolated oblique shock when its profile is measured along the free-stream direction \cite{BondarShoevTimokhin2024}. Significantly, a similar scaling is observed here for the combined transition structure along the plane of symmetry, where the incident and reflected shock-wave fronts interact. With only three incidence angles in each series, this relationship should be interpreted as a scaling observed over the investigated range rather than as a universal law.

The normalized temperature profiles in figure~\ref{fig_shock}(b) also extend over a greater streamwise distance as $\alpha$ decreases. Unlike the monotonic density transition, the temperature profiles may exhibit an overshoot above the downstream equilibrium level: $T_N>1$, corresponding to $T>T_3$. This overshoot is small for $M_n=2$, becomes clearly visible for $M_n=4$, and is particularly pronounced for $M_n=8$. Thus, decreasing $\alpha$ primarily increases the streamwise extent of the temperature transition, whereas increasing $M_n$ enhances the peak temperature
excess relative to the equilibrium value in state 3.

A similar pattern appears in the streamwise velocity, figure~\ref{fig_shock}(c). Decreasing $\alpha$ broadens the region of flow deceleration along the plane of symmetry. At high $M_n$, a local velocity minimum occurs near the reflection region, with $v_{x,N}<0$, or equivalently $v_x<v_{x3}$. This velocity undershoot is most pronounced at high $M_n$ and large incidence angles. As $\alpha$ decreases, the minimum becomes less localized and is spread over a longer region, together with the overall broadening of the profile.

Taken together, the profiles in figure~\ref{fig_shock} show that changing $\alpha$ and $M_n$ affects the flow structure in different ways. At fixed $M_n$, decreasing $\alpha$ primarily increases the extent of the transition region in units of $\lambda_\infty$. By contrast, varying $M_n$ has a stronger influence on the amplitudes of non-equilibrium features, including the temperature overshoot and the streamwise-velocity minimum relative to equilibrium state 3.

The simultaneous occurrence of a temperature overshoot and a local streamwise velocity below $v_{x3}$ indicates a pronounced non-equilibrium rearrangement of translational motion in the RR region. To characterize this process in more detail, we next consider how translational energy is distributed among the coordinate directions.

\subsubsection{Directional translational temperatures along the plane of symmetry}
\label{subsubsec_symmetry_plane_Txyz}

To examine the kinetic mechanism underlying the local maximum of the total translational temperature, we consider how translational energy is distributed among the coordinate directions. Three directional translational temperatures, $T_x$, $T_y$ and $T_z$, are introduced. The total translational temperature is defined as
\begin{equation}
    T=
    \frac{1}{3}
    \left(
        T_x+T_y+T_z
    \right).
    \label{eq_total_temperature}
\end{equation}
Here $T_x$, $T_y$ and $T_z$ are the kinetic temperatures defined by the mean energy of thermal molecular motion in the $x$, $y$ and $z$ directions, respectively. These temperatures are related to the diagonal components of the stress tensor, in the nondimensional form adopted in Eq.~\eqref{eq_Nondimensionalization}, as follows:
\begin{equation}
T_x = \frac{2\left(p+\sigma_{xx}\right)}{\rho},
\quad
T_y = \frac{2\left(p+\sigma_{yy}\right)}{\rho},
\quad
T_z= \frac{2\left(p+\sigma_{zz}\right)}{\rho}.
\label{eq_Ti}    
\end{equation} 
Here $\sigma_{ij}=P_{ij}-p\delta_{ij}$ denotes the non-isotropic part of the kinetic pressure tensor $P_{ij}$, with $\delta_{ij}$ being the Kronecker delta. This sign convention is also used in the total-energy balance below.

As with the total temperature, the directional temperatures are normalized by the full temperature change between equilibrium states 1 and 3 to enable direct comparison of cases with different $M_n$:
\begin{equation}
    T_{i,N}
    =
    \frac{T_i-T_1}{T_3-T_1},
    \qquad
    i=x,y,z.
    \label{eq_Ti_normalized}
\end{equation}
Thus, $T_{x,N}=T_{y,N}=T_{z,N}=0$ in equilibrium state 1, and all three components approach unity after complete relaxation to state 3. The value of $T_3$ in (\ref{eq_Ti_normalized}) is determined separately for each case from the corresponding inviscid solution.

With this normalization, the normalized total temperature introduced above is directly related to the directional temperatures by
\begin{equation}
    T_N
    = \frac{T-T_1}{T_3-T_1}
    =
    \frac{1}{3}
    \left(
        T_{x,N}+T_{y,N}+T_{z,N}
    \right).
    \label{eq_TN_components}
\end{equation}
This relation allows direct comparison of the non-equilibrium behaviour of the directional temperatures with the scalar translational-temperature profile in figure~\ref{fig_shock}(b).

\begin{figure}[!htbp]
\centering
\RRprofileverticalpanels
  {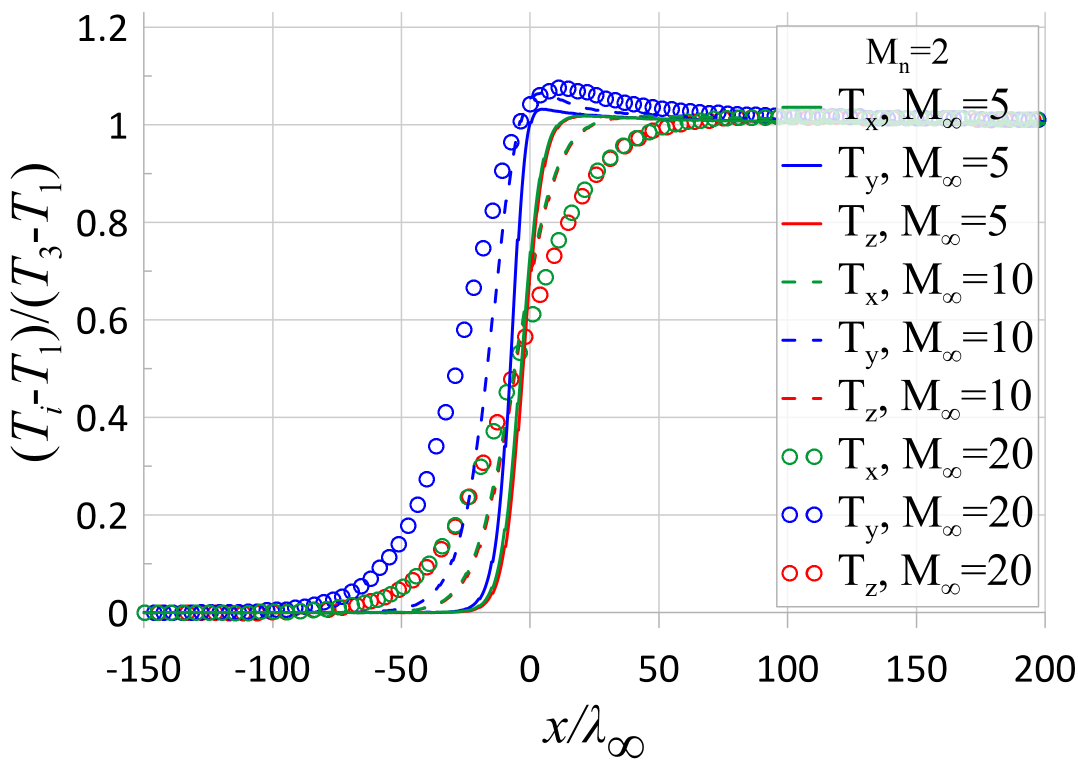}
  {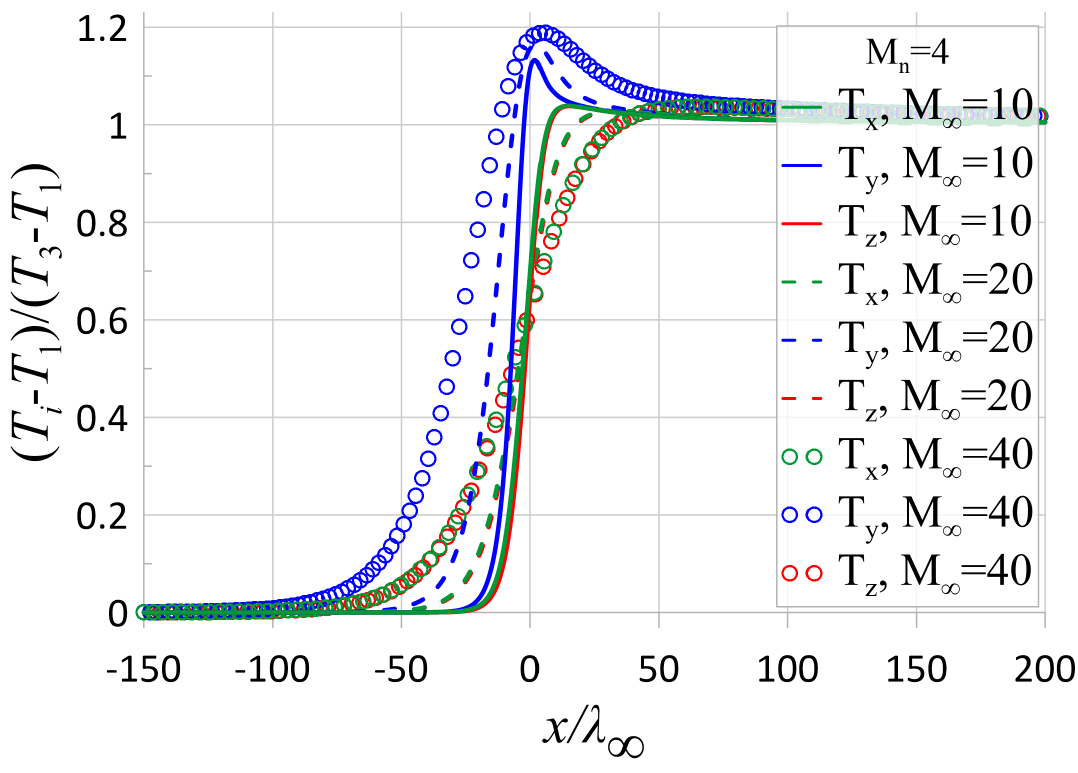}
  {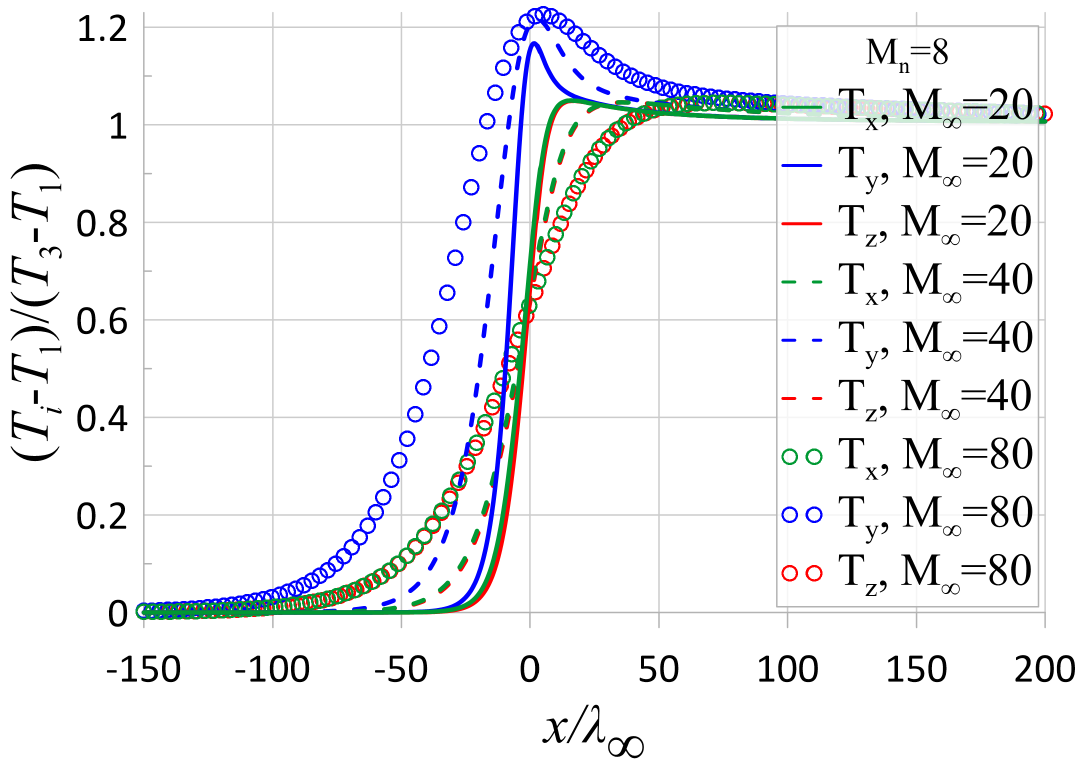}
\caption{Normalized directional translational temperatures along the plane of symmetry for
$M_n=2$ (a), $M_n=4$ (b) and $M_n=8$ (c).}
\label{fig_Ti_shock}
\end{figure}

For all nine cases, figure~\ref{fig_Ti_shock} shows a separation of directional temperatures within the transition region, directly indicating a local departure from equipartition of translational energy among the coordinate directions. The strongest deviation from equilibrium is observed in $T_y$, which corresponds to thermal molecular motion normal to the plane of symmetry.

At fixed $M_n$, decreasing the incidence angle $\alpha$ primarily increases the width of the region where the directional temperatures differ. At $\alpha=23.578^\circ$, the transition is localized to a relatively narrow region near the nominal reflection point. As $\alpha$ decreases, non-equilibrium variations begin farther upstream, and subsequent relaxation towards state 3 extends over a progressively larger distance in $x/\lambda_\infty$. This trend agrees with the broadening of the normalized density, total-temperature and streamwise-velocity profiles in figure~\ref{fig_shock}. The angle $\alpha$ also affects the amplitude of directional non-equilibrium. In all three series, $T_{y,N}$ starts to increase ahead of $T_{x,N}$ and $T_{z,N}$ and reaches a local maximum above the equilibrium level of state 3, $T_{y,N}^{\max}>1$. As $\alpha$ decreases, this maximum increases monotonically within all three series, while the region of directional non-equilibrium becomes much longer.

Comparison of the three series in figure~\ref{fig_Ti_shock} reveals a stronger dependence of the separation amplitude on $M_n$. At $M_n=2$, the maximum of $T_{y,N}$ is relatively small, although directional non-equilibrium is already clearly present. As $M_n$ increases to $4$ and then $8$, the excess of $T_y$ over the other components becomes quantitatively more obvious. Thus, the $M_n=2$ series retains the qualitative features of the mechanism, showing that they are not restricted to strong incident shocks. Increasing $M_n$ substantially amplifies them.

The increase in $T_y$ is consistent with a mechanism involving deceleration of the directed velocity component normal to the plane of symmetry and subsequent two-dimensional redistribution of translational energy in the reflection zone \cite{BondarShoevTimokhin2024}. Behind the incident oblique shock, the flow acquires a non-zero transverse velocity component $v_{2y}$ directed towards the plane of symmetry. Downstream of the reflection zone, symmetry requires $v_{y}=0$. Thus, the directed kinetic energy of transverse motion decreases substantially within the interaction region, accompanied by its redistribution among the translational degrees of freedom. During the initial stage of this process, the largest temperature increase is observed in the $y$ direction.

To compare the observed rise in $T_y$ quantitatively, consider the transverse velocity component in equilibrium state 2 immediately behind the incident shock. The oblique-shock relations give its magnitude as
\begin{equation}
    \frac{|v_{2y}|}{C_\infty}
    =
    M_\infty\sqrt{\frac{\gamma}{2}}
    \sin\alpha\cos\alpha
    \left(1-\frac{\rho_1}{\rho_2}\right)
    =
    M_n\sqrt{\frac{\gamma}{2}}
    \cos\alpha
    \left(1-\frac{\rho_1}{\rho_2}\right).
\label{eq_v2y}
\end{equation}
At fixed $M_n$, the ratio $\rho_2/\rho_1$ is constant, so the dependence of $|v_{2y}|$ on $\alpha$ within a series is determined by the factor $\cos\alpha$.

Table~\ref{tab_Vy} lists $|v_{2y}|$ and the peak overshoots of the transverse directional temperature and scalar translational temperature along the plane of symmetry. Velocity is normalized according to (\ref{eq_Nondimensionalization}). Since the signs of $v_{2y}$ are opposite in the upper and lower halves of the symmetric flow, its absolute value is used.

\begin{table}[ht]
\centering
\caption{Magnitude of the transverse velocity behind the incident shock and peak overshoots of the transverse directional and scalar translational temperatures along the plane of symmetry (DSMC values are rounded for presentation).}
\label{tab_Vy}
\begin{tabular}{rrrrrr}
\toprule
\multicolumn{1}{c}{$M_n$} &
\multicolumn{1}{c}{$M_\infty$} &
\multicolumn{1}{c}{$\alpha$, $^\circ$} &
\multicolumn{1}{c}{$|\hat v_{2y}|$} &
\multicolumn{1}{c}{$T_{y,N}^{\max}-1$} &
\multicolumn{1}{c}{$T_N^{\max}-1$} \\
\midrule
2.0 &  5.0 & 23.578 & 0.941 & 0.032 & 0.019 \\
2.0 & 10.0 & 11.537 & 1.006 & 0.065 & 0.019 \\
2.0 & 20.0 &  5.739 & 1.022 & 0.076 & 0.017 \\
\midrule
4.0 & 10.0 & 23.578 & 2.353 & 0.133 & 0.042 \\
4.0 & 20.0 & 11.537 & 2.516 & 0.163 & 0.031 \\
4.0 & 40.0 &  5.739 & 2.555 & 0.189 & 0.041 \\
\midrule
8.0 & 20.0 & 23.578 & 4.942 & 0.167 & 0.054 \\
8.0 & 40.0 & 11.537 & 5.283 & 0.218 & 0.051 \\
8.0 & 80.0 &  5.739 & 5.365 & 0.226 & 0.050 \\
\bottomrule
\end{tabular}
\end{table}

At fixed $M_n$, the magnitude $|v_{2y}|$ increases monotonically as $\alpha$ decreases. In each series, reducing $\alpha$ from $23.578^\circ$ to $5.739^\circ$ increases $|\hat v_{2y}|$ by approximately $8$--$9\%$. At the same time, the maximum overshoot of the transverse directional temperature, $T_{y,N}^{\max}-1$, increases substantially: from $0.032$ to $0.076$ for $M_n=2$, from $0.133$ to $0.189$ for $M_n=4$, and from $0.167$ to $0.226$ for $M_n=8$. Thus, the rise in $T_{y,N}^{\max}-1$ as $\alpha$ decreases is consistent with the concurrent increase in the directed transverse velocity behind the incident shock. However, the relative increase in $T_{y,N}^{\max}-1$ is much larger than the change in $|v_{2y}|$; the latter alone cannot quantitatively account for the effect. Moreover, this relationship is not a direct local energy balance, since directional temperatures are determined by two-dimensional transport and relaxation throughout the reflection zone.

The peak overshoot of the scalar translational temperature, $T_N^{\max}-1$, does not increase systematically as $\alpha$ decreases. Within each series, its variations are much smaller than the growth of $T_{y,N}^{\max}-1$; small differences between nearby extrema should not be interpreted as an established monotonic or non-monotonic trend without an estimate of statistical uncertainty. The clearest effect is therefore an increase in directional translational non-equilibrium, rather than in the scalar temperature maximum.

According to (\ref{eq_TN_components}), the scalar translational temperature is determined jointly by all three directional components. In general, the maxima of $T_y$ and $T$ occur at different locations along the plane of symmetry, so $T_{y,N}^{\max}-1$ and $T_N^{\max}-1$ cannot be interpreted directly as contributions of one quantity to the other. Nevertheless, $T_y$ exhibits the largest overshoot above the downstream equilibrium temperature $T_3$. The magnitude of this overshoot increases systematically as $\alpha$ decreases within each series and, overall, as $M_n$ increases.

Thus, the directional-temperature analysis shows that changing the reflection parameters affects 
the redistribution of translational energy among the coordinate directions.
Directional temperatures alone, however, do not characterize changes in the sum of internal and directed kinetic energies along the plane of symmetry. We therefore next consider total enthalpy to examine the combined variations in static enthalpy and directed kinetic energy.

\subsubsection{Total-enthalpy distribution along the plane of symmetry}
\label{subsubsec_symmetry_plane_Enthalpy}

The total enthalpy $H=h+\frac{v^2}{2}$ is the sum of the static enthalpy and the directed kinetic energy per unit mass. For a perfect monatomic gas,
\begin{equation}
    h=c_pT=\frac{5}{2}RT.
\end{equation}

In accordance with (\ref{eq_Nondimensionalization}), enthalpy is normalized by $C_\infty^2=2RT_\infty$. Omitting the hat notation for dimensionless quantities, we obtain
\begin{equation}
    h=\frac{5}{4}T,
    \qquad
    H=
    \frac{5}{4}T
    +
    \frac{v_x^2+v_y^2}{2}.
    \label{eq_H_nondim}
\end{equation}
For the free stream,
\begin{equation}
    H_\infty
    =
    \frac{5}{4}
    +
    \frac{1}{2}M_\infty^2\frac{\gamma}{2}
    =
    \frac{5(M_\infty^2+3)}{12}.
    \label{eq_Hinf}
\end{equation}

In inviscid adiabatic flow, total enthalpy is constant along a streamline and takes the same value on either side of a stationary shock. We therefore consider its dimensionless deviation from the free-stream value,
\begin{equation}
    \Delta H=H-H_\infty.
    \label{eq_DeltaH}
\end{equation}
In the downstream equilibrium state 3, total enthalpy recovers its free-stream value, so $\Delta H\rightarrow0$.

\begin{figure}[!htbp]
\centering
\RRprofileverticalpanels
  {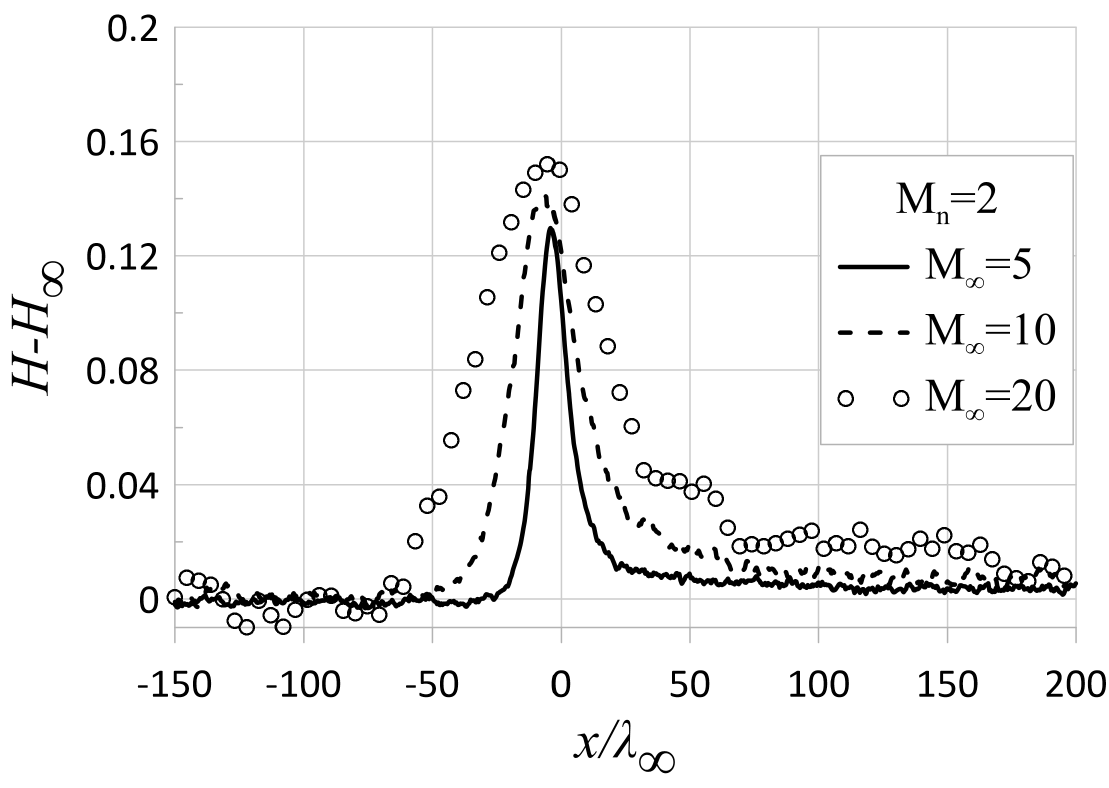}
  {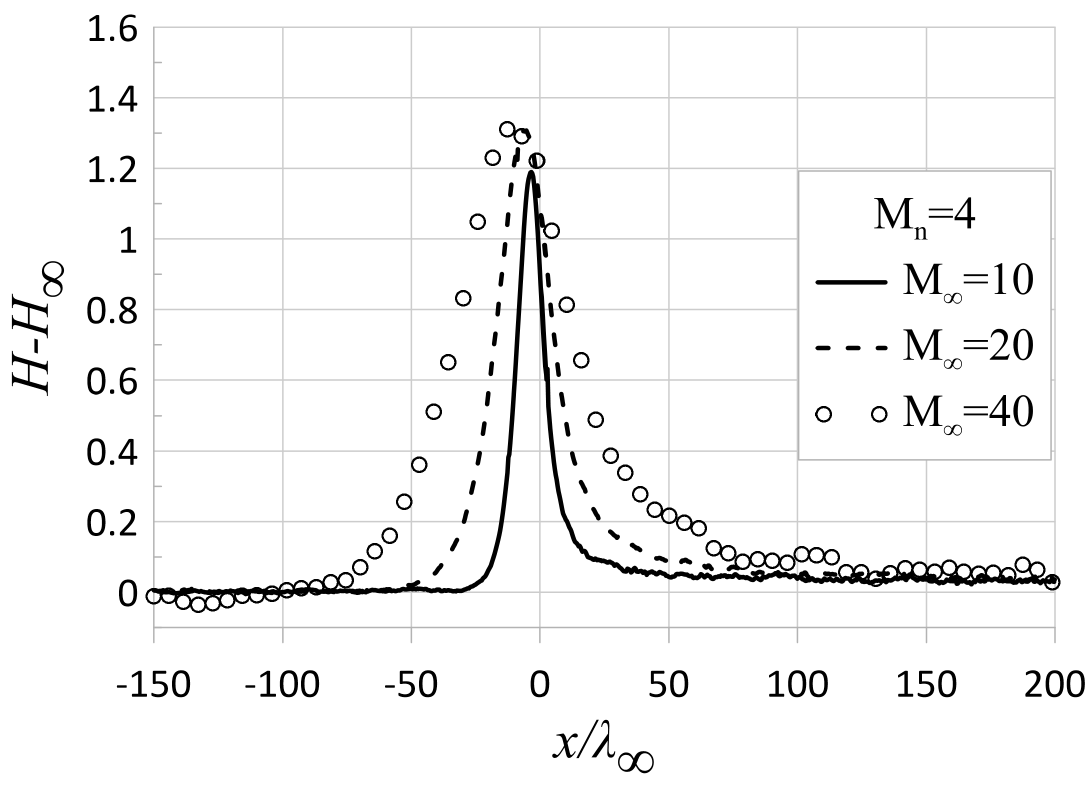}
  {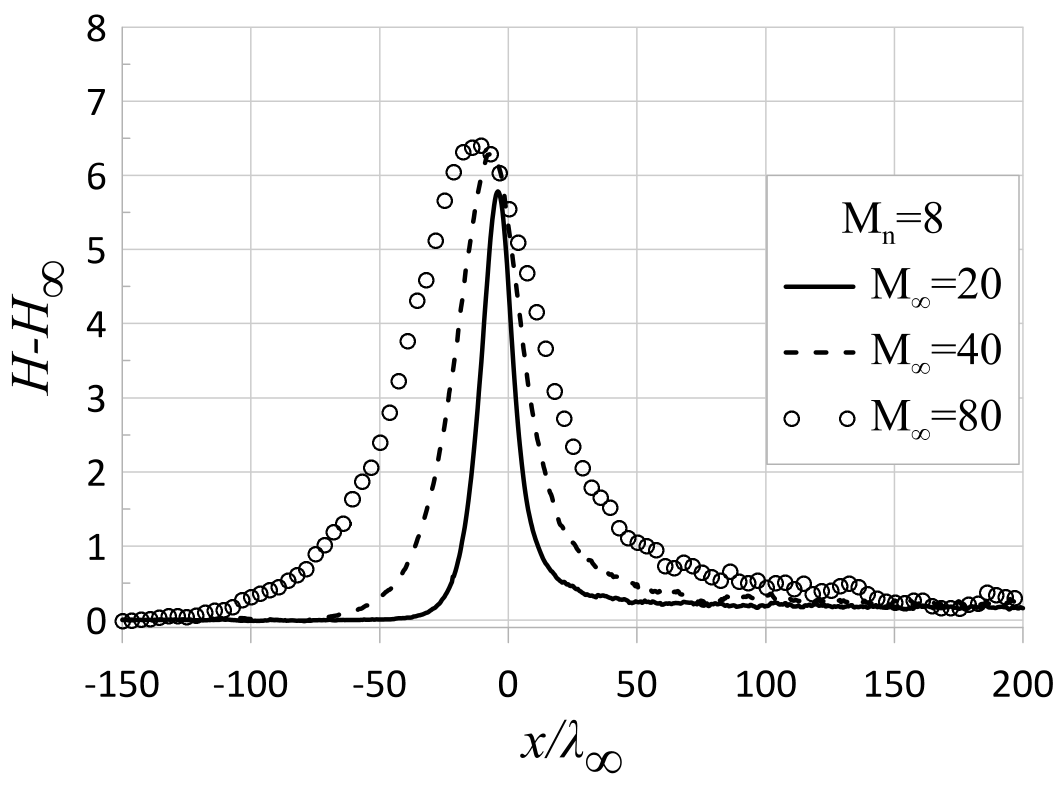}
\caption{$H-H_\infty$ along the symmetry plane
for $M_n=2$ (a), $M_n=4$ (b), and $M_n=8$ (c).}
\label{fig_enthalpy}
\end{figure}

Figure~\ref{fig_enthalpy} shows the distributions of $\Delta H$ along the plane of symmetry for all nine cases. A positive local maximum of $\Delta H$, representing a total-enthalpy excess above the equilibrium value, develops near the RR region for every set of parameters. The effects of $M_n$ and $\alpha$ on the total-enthalpy distribution are, however, different.

The peak excess depends most strongly on the incident-shock normal Mach number. The maximum of $\Delta H$ is of order $0.1$ for the $M_n=2$ series, of order unity for $M_n=4$, and reaches several units for $M_n=8$. Thus, increasing $M_n$ sharply enhances the local total-enthalpy excess in the reflection region. This trend is consistent with the stronger compression, temperature overshoot and directional-temperature anisotropy observed at larger $M_n$.

Within each series, the incidence angle has a much stronger effect on the spatial extent of the distribution than on its peak value. At the largest angle, $\alpha=23.578^\circ$, the maximum of $\Delta H$ is concentrated within a relatively narrow region near the nominal reflection point. As $\alpha$ decreases, the profile broadens both upstream and downstream, and the region over which $\Delta H$ differs appreciably from zero becomes substantially longer along the plane of symmetry.

At the smallest angle considered, $\alpha=5.739^\circ$, the deviation of $H$ from $H_\infty$ becomes noticeable much farther upstream, while $\Delta H$ returns towards zero more gradually downstream of its maximum, when measured in $x/\lambda_\infty$. Similar streamwise broadening was previously observed for the normalized density, temperature, streamwise-velocity and directional-temperature profiles. The extended total-enthalpy distribution is therefore another manifestation of the geometric rearrangement of the RR region as the incidence angle decreases.

The quantity $\Delta H$ should not be interpreted as a direct measure of the conversion of directed kinetic energy into internal energy. Local deviations of $H$ from $H_\infty$ within a transition region may result from viscous and heat-conduction transport and are not, by themselves, exclusively two-dimensional effects \cite{Enthalpy2020}.

A detailed decomposition of the energy balance for RR was presented in \cite{BondarShoevTimokhin2024}. In particular, two-dimensional terms associated with transverse compression and heat transfer in the $y$ direction were found to play an important role near the plane of symmetry, and their effects persist into the downstream wake beyond the immediate shock-interaction region.

The present $\Delta H$ distributions are consistent with this picture. The extended relaxation of $H$ towards $H_\infty$ indicates that energy redistribution does not end locally where the incident and reflected shock-wave fronts overlap. As $\alpha$ decreases, this relaxation region becomes substantially longer along the plane of symmetry, consistent with the overall spatial changes observed in density, temperature and velocity.

\section{Conclusions}
\label{sec_conclusion}

A parametric DSMC study has been conducted of the non-equilibrium region formed by the steady regular reflection of two symmetric oblique shock waves. Nine flow configurations were considered in three series at fixed incident-shock normal Mach numbers $M_n=2$, $4$ and $8$. Within each series, the incidence angle $\alpha$ was decreased from $23.578^\circ$ to $5.739^\circ$ while preserving $M_n=M_\infty\sin\alpha$. 
As $\alpha$ decreases within a series, the thermodynamic jumps across the incident shock remain constant, whereas the Mach number $M_2$ ahead of the reflected shock increases considerably and the required reverse flow deflection decreases.

The two-dimensional DSMC fields show that all nine configurations retain a two-shock RR structure. Decreasing $\alpha$ makes the interaction region of the incident and reflected shock-wave fronts longer along the plane of symmetry  and simultaneously narrows it in the transverse direction. This behaviour occurs already at $M_n=2$ and persists as the incident-shock normal strength increases to $M_n=4$ and $8$.

To compare the transition profiles directly, the density, scalar translational temperature and streamwise velocity along the plane of symmetry were normalized by their full changes between the respective equilibrium states 1 and 3. The normalized profiles exhibit a clear grouping by incidence angle. At fixed $\alpha$, profiles for different $M_n$ remain relatively close, whereas decreasing $\alpha$ at fixed $M_n$ produces a substantial increase in their  extent.
The density-transition width $L_\rho$ was introduced to quantify this effect. At fixed $M_n$ and $\omega$, the results over the investigated parameter range are consistent with the approximate scaling $L_\rho\propto1/\sin\alpha$. Whereas geometric stretching of the profile of an isolated oblique shock is expected, a similar relationship is found here for the combined transition structure of RR. This conclusion applies to the investigated ranges of incidence angle and incident-shock strength.

Increasing $M_n$ has a much stronger effect on the amplitudes of local non-equilibrium features. At comparable incidence angles, higher $M_n$ leads to a stronger temperature overshoot, a deeper local streamwise-velocity minimum relative to equilibrium state 3, and a greater separation of directional translational temperatures. The transverse directional temperature $T_y$ rises ahead of the other components and exhibits the largest overshoot above the downstream equilibrium level. The increase in $T_y$ is consistent with the deceleration of directed transverse flow in the reflection region, but is not determined by this velocity component alone. The increase in $T_{y,N}^{\max}-1$ as $\alpha$ decreases is not accompanied by a corresponding systematic increase in the scalar temperature maximum, indicating primarily stronger directional non-equilibrium.

The total-enthalpy deviation $\Delta H=H-H_\infty$ provides an additional characterization of the energy structure. A positive local maximum of $\Delta H$ forms in the RR region in every case, and its amplitude increases sharply from $M_n=2$ to $4$ and $8$. Within each series, decreasing $\alpha$ affects the spatial shape more strongly: the $\Delta H$ profile expands both upstream and downstream and relaxes over a progressively longer streamwise distance. This confirms the general nature of the spatial rearrangement of the non-equilibrium region with decreasing incidence angle.

Overall, the chosen parameterization makes it possible to compare 
two aspects of the kinetic structure of RR while keeping the normal incident-shock strength fixed within each series. Decreasing $\alpha$ at fixed $M_n$ primarily increases the  extent of the non-equilibrium region; increasing $M_n$ enhances compression, temperature anisotropy and local energy disturbances.

\section*{Acknowledgements}
DSMC calculations were performed within  the State Assignment of the Russian Federation (Project No. 124021400040-4) using the computational resources of the Equipment Sharing Center "Mechanics" of ITAM SB RAS.  
The analysis of the theoretical and computational results was supported by the Russian Science Foundation (Grant No. 22-71-10045-P).

\bibliographystyle{unsrtnat}
\bibliography{references}
\end{document}